\documentclass[a4paper,fleqn,usenatbib]{mnras}

\usepackage[T1]{fontenc}
\usepackage{ae,aecompl}

\usepackage{graphicx}	
\usepackage{amsmath}	
\usepackage{amssymb}	
\usepackage{multicol}        
\usepackage{bm}		
\usepackage{pdflscape}	
\usepackage{hyperref}

\usepackage{float}

\title[Marked correlation functions in LOWZ galaxies]{Measurement of marked correlation functions in SDSS-III Baryon Oscillation Spectroscopic Survey using LOWZ galaxies in Data Release 12}

\author[S. Satpathy et al.]{
Siddharth Satpathy$^{1,2}$\thanks{E-mail: siddharthsatpathy@cmu.edu},
Rupert Croft$^{1,2,3,4}$,  
Shirley Ho$^{1,2,5,6,7}$ \&
Baojiu Li$^{8}$
\\
$^{1}$Department of Physics, Carnegie Mellon University, 5000 Forbes Ave., Pittsburgh, PA 15213, USA\\
$^{2}$The McWilliams Center for Cosmology, Carnegie Mellon University, 5000 Forbes Ave., Pittsburgh, PA 15213, USA \\
$^{3}$ School of Physics, The University of Melbourne, VIC 3010, Australia\\
$^{4}$ ARC Centre of Excellence for All Sky Astrophysics in 3 Dimensions (ASTRO 3D), Australia\\
$^{5}$Flatiron Institute, Center for Computational Astrophysics, NY, 10010, USA  \\
$^{6}$Lawrence Berkeley National Laboratory, 1 Cyclotron Road, Berkeley, CA 94720, USA \\
$^{7}$Departments of Physics and Astronomy, University of California, Berkeley, CA, 94720, USA  \\
$^{8}$Institute for Computational Cosmology, Department of Physics, Durham University, South Road, Durham DH1 3LE, UK
}

\date{Accepted XXX. Received YYY; in original form ZZZ}

\pubyear{2018}

\begin{document}
\label{firstpage}
\pagerange{\pageref{firstpage}--\pageref{lastpage}}
\maketitle

\begin{abstract}
Marked correlation functions, which are sensitive to the clustering of galaxies in different environments, have been proposed as constraints on modified gravity models. We present measurements of the marked correlation functions of galaxies in redshift space using 
361,761 LOWZ ($z_{\rm eff} = 0.32$) galaxies from the Sloan Digital Sky Survey III (SDSS III) Baryon Oscillation Spectroscopic Survey (BOSS) Data Release 12 (DR12) and compare them to $\Lambda$CDM+General Relativity simulations. We apply mass cuts to find the best match between the redshift space autocorrelation function of subhaloes in the simulation and  in the observations. We then compare the marked correlation functions, finding no significant evidence for deviations of the marked correlation functions of LOWZ galaxies from $\Lambda$CDM on scales $6 \ h^{-1}$Mpc $ \leq s \leq$ $69 \ h^{-1}$Mpc. The constraining power of marked correlation functions in our analysis is limited by our ability to model the autocorrelation function of galaxies on small scales including the effect of redshift distortions. The statistical errors are well below the differences seen between marked correlation functions of $f$(R) gravity models and $\Lambda$CDM  in recent publications (Armijo et al., Hern\'{a}ndez-Aguayo et al.) indicating that improved future theoretical analyses should be able to rule out some models definitively.
\end{abstract}



\section{Introduction}
The observation of acceleration in the expansion of the Universe \citep[][]{Riess1998, Perlmutter1999} is one of the most important discoveries of observational cosmology. Among the many attempted explanations, those based on dark energy \citep[][]{Caldwell2002, Daly2003, Linder2003, UjjainiAlam2003, Riess2004, Copeland2006, Wood_Vasey2007, Cai2007, Buchert2008, Frieman2008}, modified theories of gravity \citep[][]{Dolgov2003, Abdalla2005, Li2007, Nojiri2007, Starobinsky2007, Bean2007, Brax2008, Faraoni2008, De_Felice2010, Jain2010, Clifton2012, Joyce2015, Huterer2015} and alternative theories which rely on inhomogeneities in the distribution of matter in the Universe \citep[][]{Kolb2006, Haavard2006, Bene2006, Krasinski2010, Chatterjee2011, Celerier2014, Skarke2014, Mertens2016} feature prominently. The variety of different possible models motivates the need to test for and to put contraints on various theories of gravity. \citet{Planck2016Ade, Koyama2016, Jimenez2016, Joyce2016} present some important advances made in this regard. More recently, the discovery of gravitational waves has opened up remarkable new avenues for the testing of and selection between various theories of gravity. Measurement of the cosmological speed of gravity with the gravitational wave event GW170817 \citep[][]{GW170817_2017PRL, GW170817_2017ApJ} has put extremely stringent bounds on the speed of gravitational waves and has served to place constraints on many modified gravity theories \citep[][]{Baker2017, Lombriser2017, Vainio2017, Amendola2017} and models of dark energy \citep[][]{Creminelli2017, Ezquiaga2017} at both large and small scales. While the discovery of the gravitational wave events have restricted the range of acceptable modified gravity theories, one needs other tools to put addtional observational constraints on different cosmological theories including modified gravity models and general relativity (GR). Development of such tools enhances the constraining power of cosmological surveys. 

 Measurements of redshift-space-distortions  (RSD) and baryon acoustic oscillations (BAO) \citep[][]{Hoyle2002, DaAngela2005, Cabre2009, Samushia2012, Macaulay2013, Yang2014, Cuesta2016, Acacia2017} can be used to provide constraints on dark energy and modified gravity theories. In such exercises, the two-point galaxy correlation function is often used as the standard statistic to study and quantify clustering of galaxies. While, this statistic uses information about the positions of galaxies, it does not directly make use of other properties such as information about the environment around galaxies. At the same time, many models of gravity have mechanisms such as screening which arise as consequences of the density around galaxies \citep[][]{Khoury2004PhRvL, Khoury2004PhRvD,  Hinterbichler2010PhRvL, Brax2010PhRvD, DavisAC2012, Brax2008}. Because of the screening mechanisms, the effects of modified gravity models are revealed in environments where the gravitational field is weak. The dependence of many modified gravity models on screening mechanisms motivates the need for a clustering estimator which includes the effect of the environments around galaxies. The proposal to test gravity models by the use of density-marked correlation functions is influenced by this requirement \citep[][]{Beisbart2000, Kerscher2000, Gottlober2002, Sheth2005, Skibba2006, White2009, White2016}.  

Recently, \citet{Valogiannis2018, Armijo2018, Hernandez-Aguayo2018} explored N-body simulations of $f(R)$ gravity models using marked correlation functions. The authors found that the differences between $f$(R) models and $\Lambda$CDM+GR universes can be captured by the marked correlation functions, where the marks are functions of the local density close to the simulated haloes. \citet{Hernandez-Aguayo2018}, show that the  galaxy marked correlation functions in certain $f$(R) models exhibit significant differences, (at the level of 1-20\%) on scales smaller than $r \leq 20$ $h^{-1}$Mpc. Apart from the fact that the main focus of our work is on observations,  our work differs from these publications  in a few ways. For example,  we work in redshift space rather than real space. Also, in \citet{Armijo2018, Hernandez-Aguayo2018}, the authors compare correlation function estimators obtained from $f$(R) models and $\Lambda$CDM universe on relatively smaller distance scales ($< 70$ $h^{-1}$Mpc), while we compute two point and marked correlation functions till significantly larger distance scales. 
 
The work discussed in this paper is based on the marked correlation function estimator proposed in \citet{White2016}. The main goal of this paper is to present measurements of the marked correlation functions computed from Sloan Digital Sky Survey galaxies at low redshifts ($z \sim 0.3$), and which can be used to constrain modified gravity models. We also aim to compare to the predictions of the $\Lambda$CDM+GR model.  Our paper is organized in the following manner. Section~\ref{sec: Data} presents details of the BOSS DR12 galaxy data set and the mock galaxy catalogues that we use in our research. Section~\ref{sec: Data} also gives a description of numerical simulations that we use to estimate the correlation functions for the theory (GR). Section~\ref{sec:Methodology} outlines the methods used for the design and computation of the two-point galaxy correlation and marked correlation functions. Section~\ref{sec:Analysis} outlines nuances of the statistical techniques used in the analysis of the correlation functions. Section~\ref{sec:Results} illustrates our results of the extent of agreement of the correlation functions obtained from observations with those computed for General Relativity. Finally, section~\ref{sec:Discussion} presents a critical analysis and summary of the results reported in section~\ref{sec:Analysis}.



\section{Data} \label{sec: Data}
In this section, we describe details of the galaxy data that we have used to find two point and marked correlation functions. For observations, we have used SDSS III BOSS DR12 LOWZ galaxies which is described in section~\ref{sec:BOSSDR12}. We use ``quick particle mesh (QPM)" mocks for covariance studies of theory and observation multipoles. Features of QPM mock catalogs are described in section~\ref{sec:MockCatalog}. Data for our theory multipoles are obtained from \textsc{Elephant} simulations \citep[][]{Cautun2017}, the details of which are outlined in section~\ref{sec:Rockstar}.

\subsection{The BOSS DR12 Galaxy Dataset} \label{sec:BOSSDR12}
In this work, we use astronomical data that were obtained by the Sloan Digital Sky Survey III (SDSS III) Baryon Oscillation Spectroscopic Survey \citep[BOSS;][]{Eisenstein2011, Dawson2013} Data Release 12 \citep[DR12;][]{Alam2015b}. The SDSS uses a dedicated 2.5m telescope at the Apache Point Observatory, New Mexico  \citep[][]{Gunn1998, Gunn2006}, equipped with two special-purpose instruments, to obtain 
spectroscopic and imaging data which covers an ensemble of galaxies, quasars, stars and ancilliary objects. The spectra for this dataset were acquired using the double-armed BOSS multifibre spectograph \citep[][]{Smee2013} and the targets selected using multicolour imaging photometry in five photometric bands \citep[$u,g,r,i,$ and $z$;][]{Fukugita1996, Gunn1998, Doi2010}. The BOSS DR12 galaxies were observed along with approximately 200,000 stars, 300,000 quasars and 400,000 ancillliary objects. The survey involved a sequence of 15-minute exposures and integration till a minimum signal-to-noise ratio was attained for the faint galaxies. This approach leads to homogeneity in the dataset with a redshift completion of over 97 per cent for the entire survey footprint. \citet{Bolton2012} discusses techniques used for
data reduction.

\begin{figure}
    \includegraphics[width=0.41\textwidth]{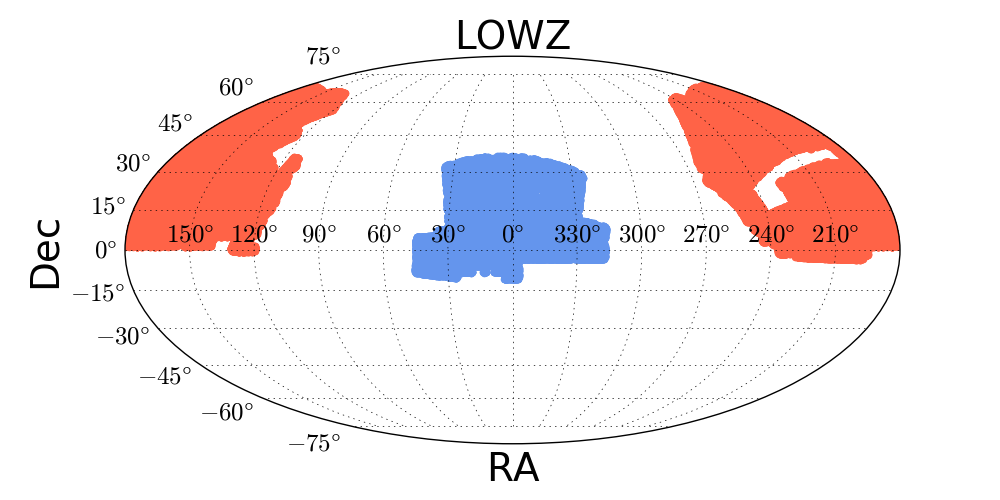}
    \includegraphics[width=0.41\textwidth]{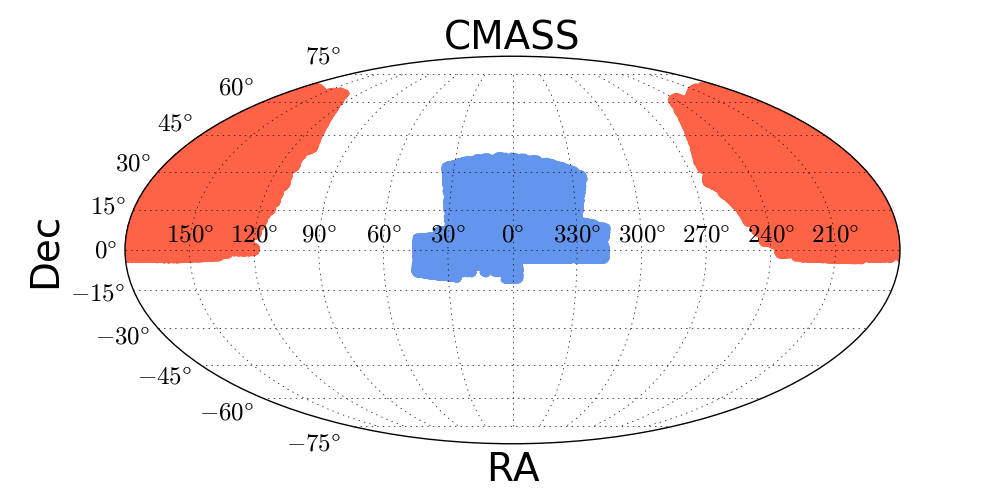}    
    \caption{A Mollweide projection of galaxies in the LOWZ and CMASS sample sets. Top: the sky coverage of galaxies in the LOWZ sample set. Bottom: the sky coverage of the CMASS galaxies. In both panels, the regions shaded red depict the North Galactic Caps while the blue regions represent the South Galactic Caps.}
    \label{fig:MockMollweide}
\end{figure}

The SDSS BOSS DR12 catalogue encompasses 1,138,964 massive galaxies partitioned into two non-overlapping redshift buckets, $viz.$ the `LOWZ' and `CMASS' redshift bins which include galaxies in the redshift ranges $0.15 < z < 0.43$ ($z_{\rm eff} = 0.32$) and $0.43 < z < 0.70$ ($z_{\rm eff} = 0.57$) respectively. \citet{Reid2016} describe methods used in the target selection of the SDSS galaxy data sets, and give details of the \textsc{mksample} code used to create large scale structure catalogues. The LOWZ DR12 sample encompasses a total of 361,762 galaxies with 248,237 galaxies in the North Galactic Cap (NGC) and 113,525 galaxies in the South Galactic Cap (SGC). The total sky coverage of the LOWZ DR12 sample is 8337.47 deg$^2$. The CMASS DR12 sample includes a total of 777,202 galaxies with 568,776 galaxies in NGC and 208,426 galaxies in SGC. The CMASS DR12 galaxies have an effective sky coverage of 9376.09 deg$^2$. Fig.~\ref{fig:MockMollweide} illustrates the relative sky coverage of NGC and SGC for the LOWZ and CMASS galaxies. In Table~\ref{tab:Sky_coverage}, we give the values of effective areas covered by NGC and SGC for the LOWZ and CMASS galaxies. 

The investigation of marked correlation functions presented in this paper is based on the use of 361,762 massive galaxies from the LOWZ redshift bin ($0.15 < z < 0.43$) of BOSS DR12 data set. In this paper, we follow the fiducial cosmology chosen in \citet{Cuesta2016}. This cosmology assumes a flat $\Lambda$CDM-GR model with Hubble constant $h \equiv H_0/(100  \ {\rm km} \ {\rm s}^{-1} \ {\rm  Mpc}^{−1})=0.70$, $\Omega_{\rm m}=0.29$, $\Omega_{\Lambda}=0.71$, $\Omega_{\rm b} h^2 = 0.02247$, $\Omega_{\rm \nu} h^2 = 0.0$, $\Omega_{\rm k}=0$ and $\sigma_8=0.8$. This selection of cosmology is inspired by Planck+BAO ($viz.$ Planck+LOWZ+CMASS+6dF+LyA) constraints in the $\Lambda$CDM model in \citet{Anderson2014}.

\begin{table}
	\centering
	\caption{Sky coverage (in deg$^2$): LOWZ and CMASS sample sets}
	\label{tab:Sky_coverage}
	\begin{tabular}{lccc}
		\hline
		\hline
		& NGC & SGC & Total \\
		\hline
		\textsc{LOWZ DR12 } & 5836.21 & 2501.26 & 8337.47 \\	
		\textsc{CMASS DR12 } & 6851.42 & 2524.67 & 9376.09  \\			
		\hline
	\end{tabular}
\end{table}

\subsection{Low resolution simulations and mock galaxy catalogues} \label{sec:MockCatalog}
A comprehensive analysis of the BOSS DR12 galaxy data sets requires the use of mock galaxy catalogues. In our work, we use mock galaxy catalogues which are generated by the use of the ``quick particle mesh" (QPM) method \citep[][]{White2014}. In the quick particle mesh method, many large volume, approximate mock catalogues are created at low computational cost by making use of low-resolution particle mesh simulations to generate the large scale dark matter density field. These particle mesh simulations are based on a flat $\Lambda$CDM cosmology with $h=0.7$, $\Omega_{\rm m}=0.274$, $\Omega_{\rm b} = 0.046$, $\Omega_{\Lambda} = 0.726$, $n=0.95$ and $\sigma_8 = 0.8$. The high resolution N-body simulations which form the foundation of the QPM mocks are based on the aforementioned cosmology. These high-resolution simulations use the TreePM$^2$ code outlined in \citet{White2002} where $3000^3$ particles ($5.9 \times 10^{10}$ $h^{-1}M_{\odot}$) are evolved in a box of side 2750 $h^{-1}$Mpc. Details of the aforesaid N-body simulations can also be found in \citet{White2011, White2012, ReidWhite2011}. 

In a manner similar to the division of the BOSS DR12 data into two redsift bins ($z_{\rm eff} = 0.32$ and $z_{\rm eff} = 0.57$), the QPM mock catalogues are partitioned into two redshift bins, $0.15 < z < 0.43$ and $0.43 < z < 0.70$ with effective redshifts of $z_{\rm eff} = 0.32$ and $z_{\rm eff} = 0.57$ respectively. For the analysis presented in this paper, we have used 100 mocks from the redshift bin $0.15 < z < 0.43$. This choice of QPM mocks is in accordance with our choice of galaxies from the LOWZ redshift bin of the SDSS BOSS DR12 catalogue. These chosen mock universes mimic the observations from SDSS galaxies and enable us to compute robust error estimates for the correlation functions obtained from the LOWZ galaxies.

\subsection{Numerical simulations of modified gravity} \label{sec:Rockstar}
For study of theoretical clustering of galaxies in $\Lambda$CDM universe, we use the \textsc{elephant} (\textsc{E}xtended \textsc{LE}nsing \textsc{PH}ysics using {AN}alytic ray \textsc{T}racing) numerical simulations which are implemented using the \textsc{ecosmog} code \citep[][]{Li2012}. The \textsc{ecosmog} code is built using the adaptive mesh refining (\textsc{amr}) N-body code \textsc{ramses} \citep[][]{Teyssier2002}. These simulations are based on boxes of size $L_{\rm box}=1024 \ h^{-1}$Mpc with $N_p=1024^3$ DM particles with a mass of $m_{\rm p}=7.798 \times 10^{10} h^{-1}M_{\odot}$. The simulations assume a flat $\Lambda$CDM cosmology with $h=0.697$, $\Omega_{\rm m}=0.281$, $\Omega_{\rm b} = 0.046$ and $\Omega_{\Lambda} = 0.719$. The said cosmological parameters follow the best-fitting values to WMAP 9 year CMB measurements \citep[][]{Hinshaw2013}. These simulations have 37+1 snapshots which were generated using initial conditions produced at $z_{\rm ini}=49$ by the \textsc{MPgrafic} code \citep[][]{Prunet2008}. Also, these simulations incorporate GR and several modified gravity models. The halo catalogues for these simulations were generated using the \textsc{rockstar} halo finder code \citep[][]{Behroozi2013}. The \textsc{rockstar} code uses particles and substructures in the halo and the spherical overdensity method described in \citet{Cole1996} to compute halo masses. 

For the purpose of our project, we consider GR catalogues from the \textsc{elephant} simulations which include both main haloes and subhaloes. Hereafter, we shall refer to these as `subhaloes' throughout the paper. These subhaloes are at a snapshot at an effective redshift of $z_{\rm eff} = 0.3470$. This dataset comprises  1,902,278 massive subhaloes with a mean mass of $7.18 \times 10^{12} \ h^{-1}M_{\odot}$. The effective redshift of these GR simulation subhaloes ($z_{\rm eff} = 0.3470$) is close to the effective redshift of LOWZ galaxies ($z_{\rm eff} = 0.32$). Consequently, we use results from this simulation as the theory to which we compare results obtained from LOWZ galaxies.

\begin{figure*}
\begin{multicols}{2}
    \includegraphics[width=0.85\linewidth]{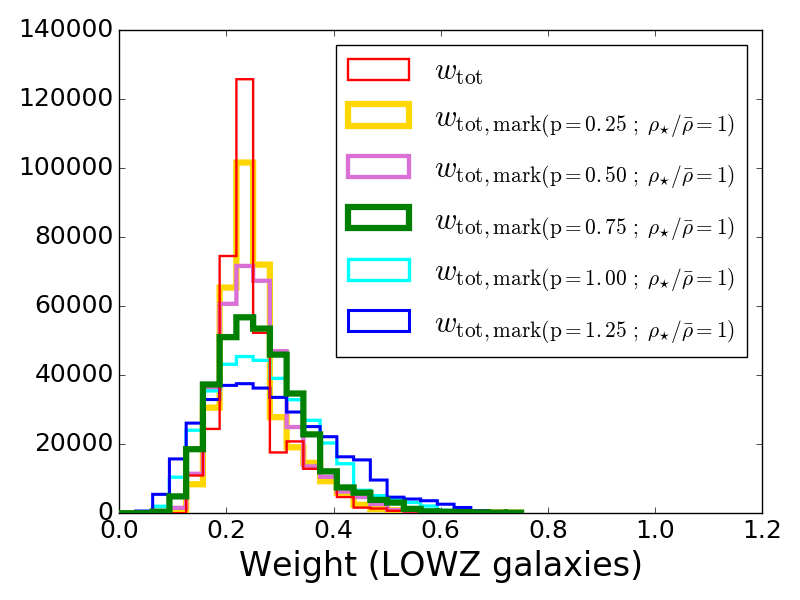}\par 
    \includegraphics[width=0.85\linewidth]{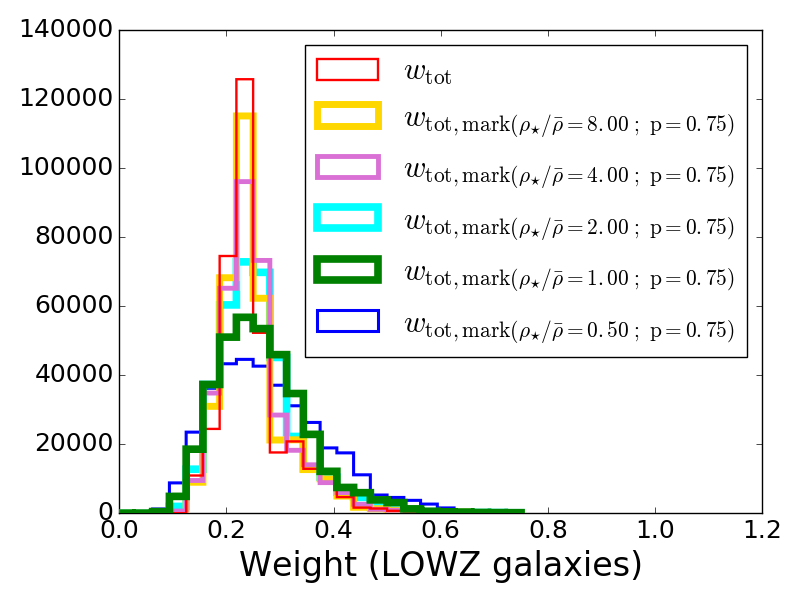}\par 
\end{multicols}
\caption{The distributions of `marked total weights', $viz.$ $w_{\rm tot, mark}$  for BOSS DR12 LOWZ galaxies for different values of parameters $p$ and $\rho_{ \star }$. The left plot shows how the distributions of $w_{\rm tot, mark}$ change when one varies the parameter $p$ in the mark $m$ in equation~\ref{eqn:mark} while keeping the ratio between the free parameter $\rho_{ \star }$ and the mean density $\bar{\rho}$ constant. In the left plot, we fix the ratio $\rho_{ \star } / \bar{\rho} = 1$ while we vary $p$ between the values 0.25 and 1.25. The trends in the change of the histograms that we see (as one changes the values of $p$ in $w_{\rm tot, mark}$) is as per expectations from the analytic form of equation~\ref{eqn:mark}. In the plot in the right, we show variation in histograms of $w_{\rm tot, mark}$ for LOWZ galaxies, when we fix the parameter $p$ in the mark $m$ and vary the ratio $\rho_{ \star } / \bar{\rho}$. Specifically, in the histograms plotted in the figure on the right, we have fixed $p=0.75$ in the mark $m$ in equation~\ref{eqn:mark} while varying the ratio $\rho_{ \star } / \bar{\rho}$ between 0.5 and 8.0. The variations in the histograms and the trends that we see when vary the ratio $\rho_{ \star } / \bar{\rho}$ from small values of 0.5 to higher values of 0.8 is as per expectations from the analytic form of equation~\ref{eqn:mark}.}
\label{fig:LOWZ_MarkWts}
\end{figure*}

\begin{figure}
    \includegraphics[width=0.41\textwidth]{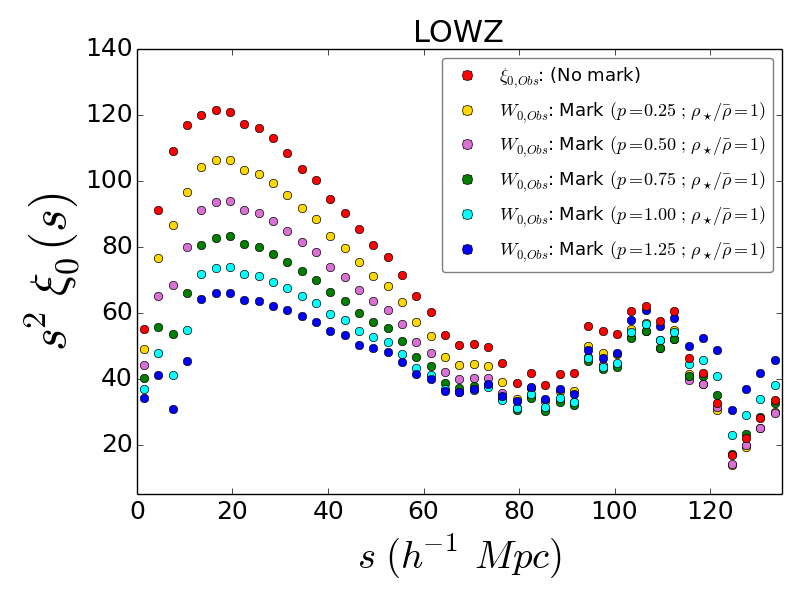}
    \includegraphics[width=0.41\textwidth]{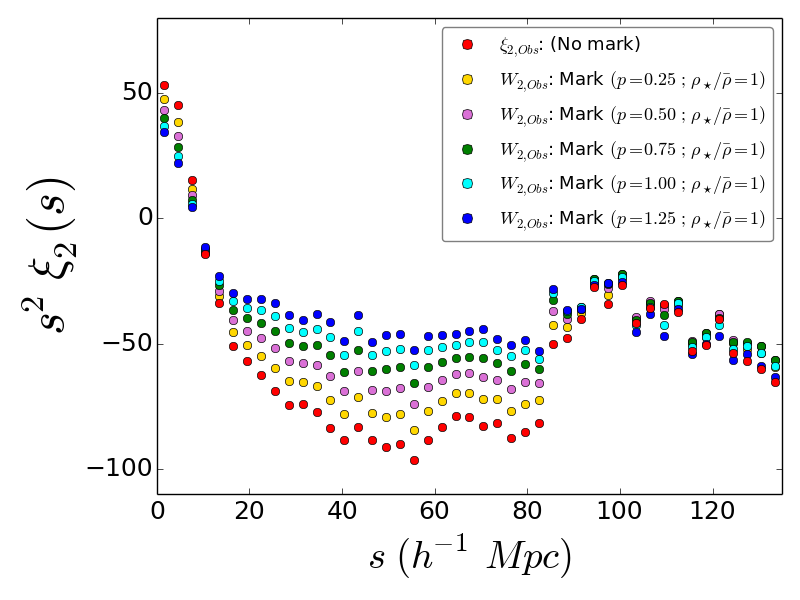}
    \caption{The top and the bottom panels represent the variations in monopole ($\xi_0(s)$) and quadrupole ($\xi_2(s)$) in LOWZ galaxies when different weights, $viz.$ $w_{\rm tot, mark}$ are used.  More specifically, these plots illustrate the variations in multipoles of LOWZ galaxies when one varies the parameter $p$ in the mark $m$ from equation~\ref{eqn:mark} (while fixing $\rho_{ \star } / \bar{\rho} = 1$).  Histograms of weights corresponding to these correlation functions are shown in the left plot of Fig.~\ref{fig:LOWZ_MarkWts}. For all plots shown here, we show data points for bins with sizes of $\Delta s = 3 h^{-1}$Mpc. The sharp changes that are seen in the correlation function multipoles around $s=10 \ h^{-1}$Mpc can be attributed to the value of local radius ($s=10  \ h^{-1}$Mpc) that we have chosen in our computation of local densities ($\rho_s$) around galaxies (using equation~\ref{eqn:rho}).}   
    \label{fig:LOWZ_Corr_Diff_p}     
\end{figure} 

\begin{figure}
    \includegraphics[width=0.41\textwidth]{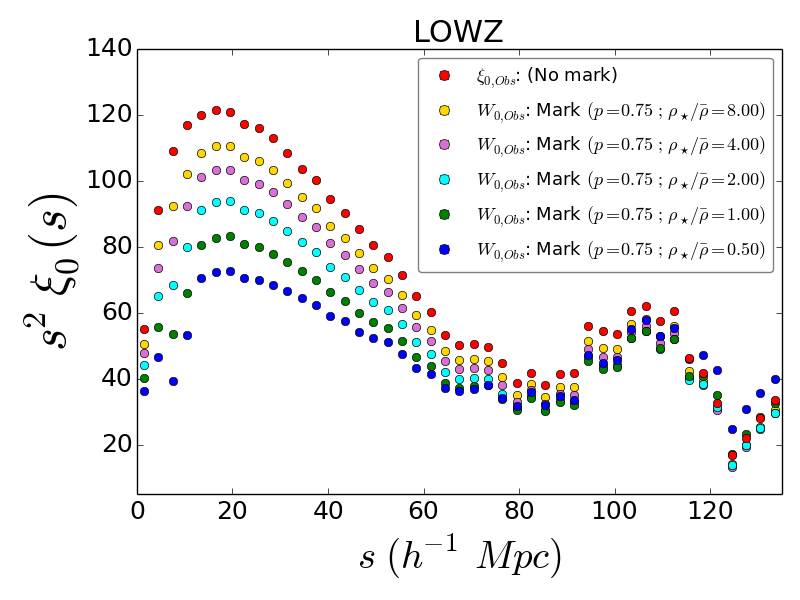}
    \includegraphics[width=0.41\textwidth]{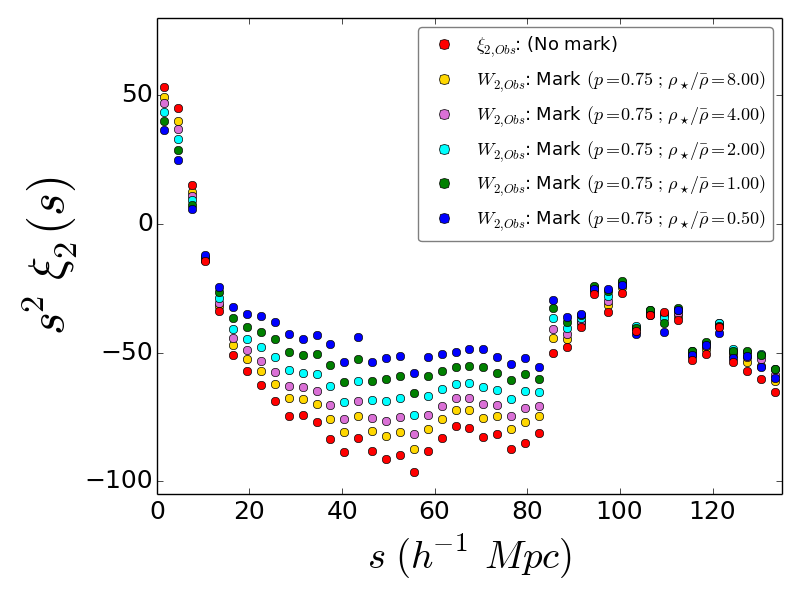}
    \caption{Monopoles and quadrupoles (in LOWZ sample) that are obtained when we vary the ratio $\rho_{ \star } / \bar{\rho}$ while computing the weights ($w_{\rm tot, mark}$). In the correlation functions shown, we use weights and marks described in equations~\ref{eqn:Anderson_mark} and~\ref{eqn:mark} respectively. The top plot depicts variations in monopoles in LOWZ galaxies when the ratio $\rho_{ \star } / \bar{\rho}$ is varied between the ranges 0.5 and 8, while the bottom plot shows variations in quadrupoles when the ratio $\rho_{ \star } / \bar{\rho}$ is varied in the same range. For all these correlation functions, the value of $p$ in the mark $m$ is fixed at a value of 0.75. Also, in all plots shown here, we show data points for bin sizes of 3 $h^{-1}$Mpc. Histograms of the weights used to obtain these correlation functions are shown in the right plot of Fig.~\ref{fig:LOWZ_MarkWts}. As in Fig.~\ref{fig:LOWZ_Corr_Diff_p}, we observe sudden changes in multipoles around $s=10 \ h^{-1}$Mpc. These are due to our choice of local radius ($s=10  \ h^{-1}$Mpc) during the computation of local densities around galaxies in the LOWZ galaxy catalogue.}
    \label{fig:LOWZ_Corr_Diff_rhoratio}    
\end{figure}

\begin{figure}
    \includegraphics[width=0.41\textwidth]{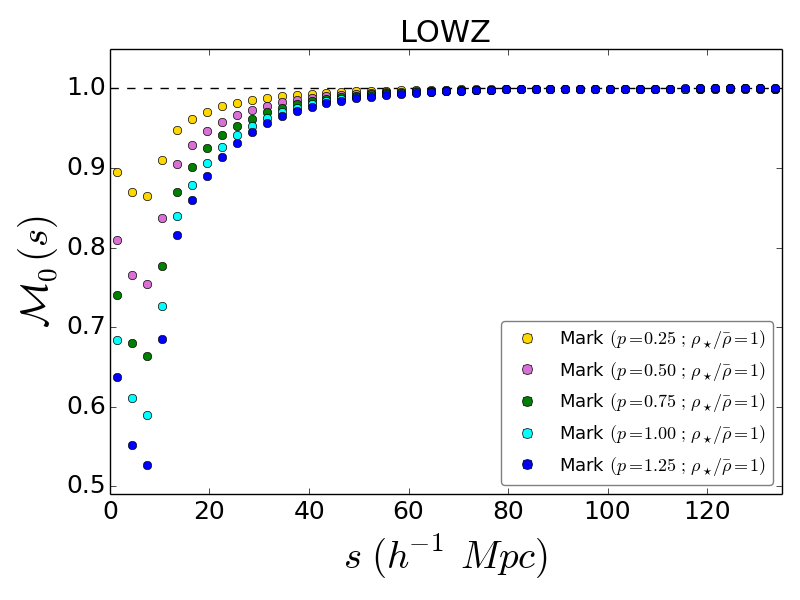}
    \includegraphics[width=0.41\textwidth]{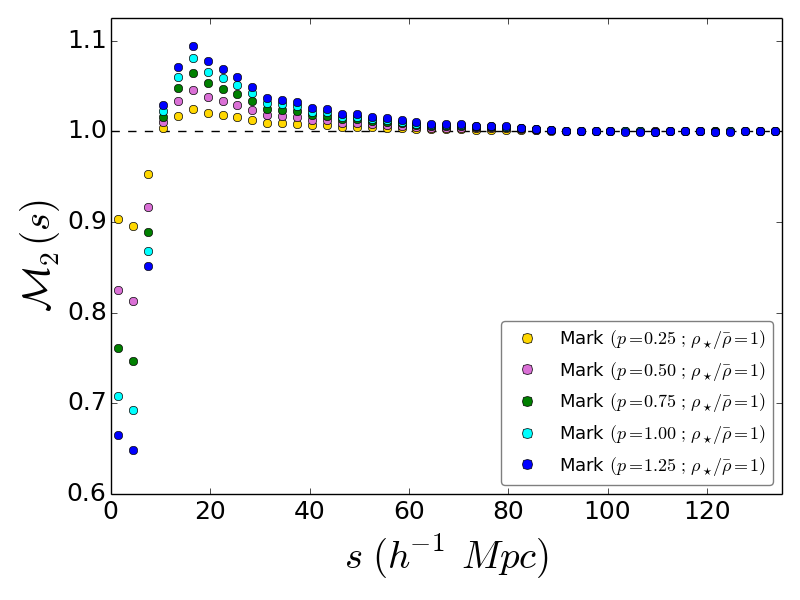}
    \caption{The marked correlation function multipoles of LOWZ galaxies. The top panel shows marked correlation function monopoles while the plots in the bottom panel show the marked correlation function quadrupoles. The different marked correlation function multipoles illustrated in these figures correspond to analysis of standard ($\xi_{0.2}$) and mark weighted ($W_{0,2}$) galaxy correlation functions obtained from LOWZ galaxy catalogue (shown in Fig.~\ref{fig:LOWZ_Corr_Diff_p}). Specifically, these marked correlation functions show variations that result when the value of $p$ in mark $m$ (equation~\ref{eqn:mark}) is varied between 0.25 and 1.25. In all the plots shown here, a constant value of the ratio $\rho_{ \star } / \bar{\rho}$ is maintained ($\rho_{ \star } / \bar{\rho}=1$). The sudden fluctuation in values of marked correlation functions that are seen near $s=10 \ h^{-1}$Mpc can be attributed to our choice of local radius ($s=10  \ h^{-1}$Mpc) in the computation of local densities $\rho_s$ using equation~\ref{eqn:rho}.}
    \label{fig:LOWZ_Mark_Diff_p}    
\end{figure} 

\begin{figure}
    \includegraphics[width=0.41\textwidth]{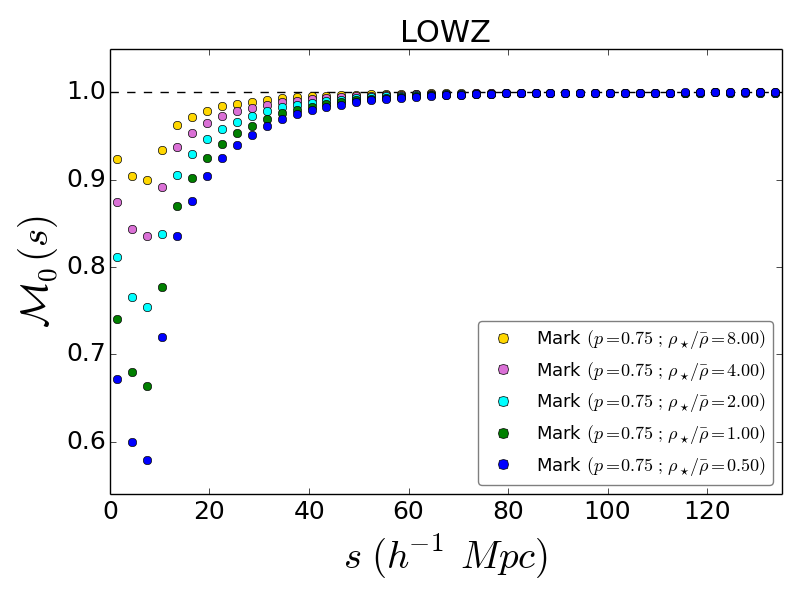}
    \includegraphics[width=0.41\textwidth]{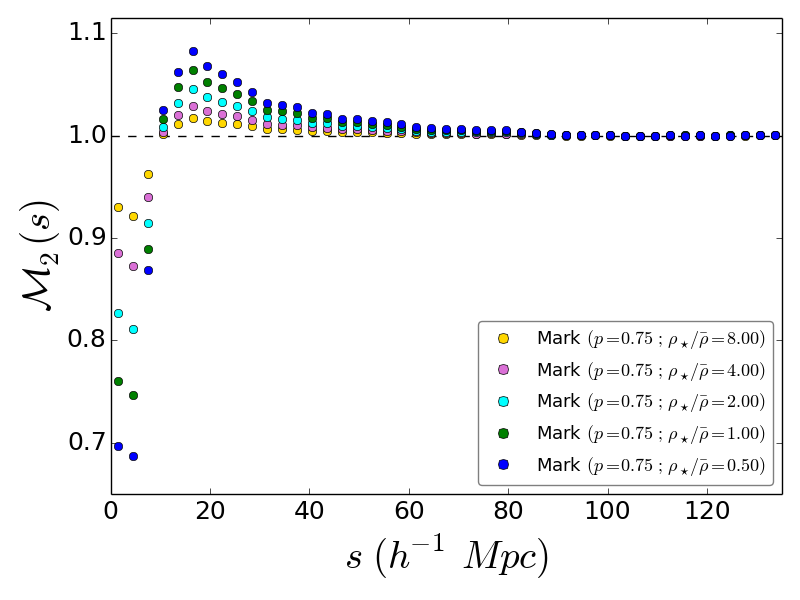}
    \caption{The various multipoles for the marked correlation functions of LOWZ galaxies which are obtained when the ratio $\rho_{ \star } / \bar{\rho}$ in the equation for mark $m$ (\textit{i.e.} equation~\ref{eqn:mark}) is varied between the values 0.5 and 4. Also, the parameter $p$ in the mark $m$ is kept constant ($p=0.75$) during the computation of these marked correlation function multipoles. The isotropic standard ($\xi_{0.2}$) and mark weighted ($W_{0,2}$) multipoles which were used to obtain these marked correlation function monopoles and quadrupoles are shown in Fig.~\ref{fig:LOWZ_Corr_Diff_rhoratio}. Similar to our observations in Fig.~\ref{fig:LOWZ_Mark_Diff_p}, we see changes in values of marked correlation function multipoles around $s=10 \ h^{-1}$Mpc. These changes can be attributed to our selection of local radius ($s=10  \ h^{-1}$Mpc) in the computation of local densities ($\rho_s$) around galaxies.}    
    \label{fig:LOWZ_Mark_Diff_rhoratio}
\end{figure}

\section{Methodology} \label{sec:Methodology}

In this section, we sketch the various techniques that we have used for the analysis of the BOSS DR12 galaxies and the QPM mocks. More exactly, in section~\ref{sec:TwoPCorrFunc} we describe the method used to obtain two point galaxy correlation function multipoles ($\xi_0(s)$ and $\xi_2(s)$). In section~\ref{sec:MarkWtCorrFunc}, we explain the procedure used to obtain weights for `mark weighted correlation functions' ($W_0(s)$ and $W_2(s)$). And, in section~\ref{sec:MarkCorrFunc}, we outline the technique used to obatin `marked correlation function multpoles' ($\mathcal{M}_0(s)$ and $\mathcal{M}_2(s)$). Throughout this paper, we use the symbol $s$ to denote distances (pair separations) in redshift space.

\subsection{The two-point galaxy correlation function} \label{sec:TwoPCorrFunc}
We use the Landy-Szalay estimator \citep[][]{Landy1993} to obtain two-point correlation functions for LOWZ galaxies, QPM mocks and galaxies in the GR simulation catalogue. The Landy-Szalay estimator is commonly used to generate and investigate anistropic two point correlation functions in galaxy samples under study. The formula for computation of the Landy-Szalay estimator $\hat{\xi}_{\rm LS}(s, \mu)$ is given in equation~\ref{eqn:LSEqn_Anisotropic}.

\begin{equation}
\label{eqn:LSEqn_Anisotropic}
\hat{\xi}_{\rm LS}(s, \mu) = \frac{DD(s, \mu) - 2DR(s, \mu) + RR(s, \mu)}{RR(s, \mu)}.
\end{equation}

In equation~\ref{eqn:LSEqn_Anisotropic}, the letter $s$ represents the pair separation between galaxies in redshift space. Also, for any given pair of two galaxies, $\mu = {\rm cos} \ \theta$, where the symbol $\theta$ denotes the angle between the pair separation vector for the two galaxies and the mean of the line of sight vectors of those two galaxies. $DD(s, \mu = {\rm cos} \ \theta)$ depicts the number of pairs of galaxies which are at a pair separation distance $s$ and angle $\theta$. Similarly,  $RR(s, \mu = {\rm cos} \ \theta)$ represents the count of pairs of randoms which have pair separation of $s$ and angle  $\theta$. $DR(s, \mu = {\rm cos} \ \theta)$ corresponds to the number of cross-pairs between galaxies and randoms which have pair separation $s$ and angle $\theta$.

Compared to other alternatives \citep[][]{DavisPeebles1983, Hamilton1993}, the anisotropic two-dimensional two-point correlation function obtained by the use of the Landy-Szalay estimator ($\xi(s, \mu)$) delivers better performance at large scales \citep[][]{Pons1999, Kerscher2000}. 

Analysis of  the two-dimensional two-point correlation function using the Landy-Szalay estimator ($\xi(s, \mu)$) will involve many bins (because of the two dimensional nature). Such an analysis would entail the use of a large covariance matrix, which would require the creation of a huge number of mocks. Such an exercise would be computationally very expensive. Because of this, we work with isotropized versions ($\tilde{\xi}_{f}(s)$) of the two-point correlation function ($\hat{\xi}( s, \mu )$) where specific kernels ($f(s, \mu)$) are used to condense the information contained in the two-point correlation ($\hat{\xi}( s, \mu )$). Such an isotropic correlation function ($\tilde{\xi}_{f}(s)$) allows  analysis of galaxy clustering with a manageable number of bins. 

\begin{equation}
\label{eqn:Isotropized2PCorr}
\tilde{\xi}_{f}(s) = \int f(s, \mu) \hat{\xi}( s, \mu ) dV.
\end{equation}

\citet{Hamilton1993} illustrated the use of the orthonormal basis of Legendre polynomials ($P_{\ell}(\mu)$) as kernels to marginalize the two-point correlation function ($\hat{\xi}( s, \mu )$) to generate the isotropic correlation functions. We use the approach outlined in \citep[][]{Hamilton1993} to obtain isotropic correlation functions ($\tilde{\xi}_{\ell}(s)$, multipoles) of various orders, \textit{i.e.} the monopole and the quadrupole ($\ell = 0, 2$ respectively) in our analysis of the LOWZ galaxy catalogue. Equation~\ref{eqn:LSEqn_Isotropic} shows how the isotropic correlation function $\tilde{\xi}_{\ell}(s)$ can be obtained from the two-point correlation function $\hat{\xi}( s, \mu )$.

\begin{align}
\label{eqn:LSEqn_Isotropic}
\tilde{\xi}_{\ell}(s) &= \frac{2 \ell + 1}{2} \int^{1}_{-1}\hat{\xi}_{\rm LS}(s,\mu) P_{\ell}(\mu)d\mu \nonumber \\
&\approx  \frac{2 \ell + 1}{2} \displaystyle\sum_{j} \Delta \mu_j \hat{\xi}_{\rm LS}(s,\mu_j) P_{\ell}(\mu_j).
\end{align}

For the research presented in this paper, we have used 100 bins in $\mu$ in the computation of all anisotropic two-point correlation functions $\xi(s, \mu)$. Furthermore, all the multipoles ($\xi(s)$) that we compute for BOSS DR12 LOWZ galaxies, QPM mocks and catalogues from \textsc{elephant} GR simulations have evenly spaced bins of width $3 \ h^{-1}$Mpc in $s$.

\subsection{Mark weighted correlation functions} \label{sec:MarkWtCorrFunc}
As a precursor to the computation of `marked correlation functions $\mathcal{M}_{\ell}(s)$', we need to compute the `mark weighted correlation functions $W_{\ell}(s)$' in addition to the standard correlation functions $\xi_{\ell}(s)$ (which are computed with the use of the weights given in equations~\ref{eqn:Anderson}.) The computation of the mark weighted correlation functions $W_{\ell}(s)$ is the first step where the local environment around galaxies is taken into consideration. The calculation of mark weighted correlation function $W_{\ell}(s)$ is a two step process. In the first step, we compute the total galaxy weights ($w_{\rm tot}$) described in equation~\ref{eqn:Anderson} for all galaxies in a given catalogue (and $w_{\rm FKP}$ to the corresponding randoms). In the second step, we calculate weights for all galaxies in the catalogue depending on the local densities around the galaxies. We compute the local density around each galaxy from a weighted count of all galaxies inside a sphere of a given (local) radius around the given galaxy. Points from the random catalogue are also used in the estimation. Specifically, we compute the local density ($\rho_s$) around a given galaxy in the following fashion:
\begin{equation}
\rho_s =  \left( \frac{w_{\rm tot_{G, s}}}{w_{\rm tot_{R, s}}} \right) \times \left( \frac{{\rm Total}_{\rm R}}{{\rm Total}_{\rm G}} \right). 
\label{eqn:rho}
\end{equation}
where $w_{\rm tot_{G, s}}$ denotes the sum of weights of all galaxies inside a sphere of radius $s$ around a given galaxy, $w_{\rm tot_{R, s}}$ represents the sum of weights of randoms in the vicinity of the given galaxy (inside a sphere of radius $s$). ${\rm Total}_{\rm G}$ and ${\rm Total}_{\rm R}$ denote the net sums of weights of all galaxies and all randoms in the selected sample. For all the analysis presented in this paper, we have considered local radii ($s$) of 10 $h^{-1}$Mpc around all galaxies while computing the local densities around those galaxies.

Following the computation of the local density around a given galaxy, we compute the mark `$m$' of that galaxy. The mark of a given galaxy is a function of the local density around the galaxy. In our case, we obtain the mark of a given galaxy `$m$' from its local density `$\rho_s$' using one of the prescriptions suggested in \citet{White2016}:
\begin{equation}
m = \left( \frac{ \rho_{ \star } + \bar{\rho} }{\rho_{ \star } + \rho_s}  \right)^p.
\label{eqn:mark}
\end{equation}
Here $\rho_s$ is the density obtained from equation~\ref{eqn:rho}, $\bar{\rho}$ is the mean density of the Universe and $\rho_{ \star }$ and $p$ are parameters that we can adjust. The choice of the form of the weight (mark) considered in equation~\ref{eqn:mark} is inspired by our aim of detecting potential signatures of modified gravity in regions where the density of galaxies is small (\textit{i.e} regions of weak gravity).

Following our description of the nature and form of marks for galaxies in equation~\ref{eqn:mark}, we are now in a position to define the total galaxy weights for `mark weighted correlation functions'. Our definition of `marked total weights' in equation~\ref{eqn:Anderson_mark} is influenced by the definition of total galaxy weights in equation~\ref{eqn:Anderson}.

\begin{equation}
\label{eqn:Anderson_mark}
w_{\rm tot, mark} = \left[ \left( w_{\rm cp} + w_{\rm noz} - 1  \right) w_{\rm star} w_{\rm see} w_{\rm FKP} \right] \times m.
\end{equation}

The prescriptions for the construction of marks (given in equation~\ref{eqn:mark}) and weights (given in equation~\ref{eqn:Anderson_mark}) ensure that the weighted correlation functions depend on the enivornments (local densities) around galaxies. The influence of low density regions on the mark $m$ (and, consequently on $W_{\ell}(s)$) is enhanced when one chooses positive values of $p$ in equation~\ref{eqn:mark}. Fig.~\ref{fig:LOWZ_MarkWts}, illustrates histograms of marked total weights $w_{\rm tot, mark}$ for galaxies in the BOSS DR12 LOWZ catalogue for different values of the parameter $p$ and $\rho_{ \star }$ in the mark $m$.

Hereafter, we will denote anisotropic two-point correlation functions computed with the weights given in equation~\ref{eqn:Anderson_mark} and the Landy-Szalay estimator described in equation~\ref{eqn:LSEqn_Anisotropic} by the symbol $W(s,\mu)$. In the same spirit, we will denote all isotropic correlation functions using weights from equation~\ref{eqn:Anderson_mark} and marginalization given in equation~\ref{eqn:LSEqn_Isotropic} by the symbol $W_{\ell}(s)$. Fig.~\ref{fig:LOWZ_Corr_Diff_p} and~\ref{fig:LOWZ_Corr_Diff_rhoratio} show plots of multipoles ($W_{0}(s)$ and $W_{2}(s)$)  obtained when marks with different values of parameters $p$ and $\rho_{ \star }$ are used as weights for LOWZ galaxies. Histograms of distributions of weights corresponding to the parameters $p$ and $\rho_{ \star }$ used to obtain multipoles in Fig.~\ref{fig:LOWZ_Corr_Diff_p} and~\ref{fig:LOWZ_Corr_Diff_rhoratio} are shown in Fig.~\ref{fig:LOWZ_MarkWts}.

\begin{figure}
    \centering
    \includegraphics[width=0.45\textwidth]{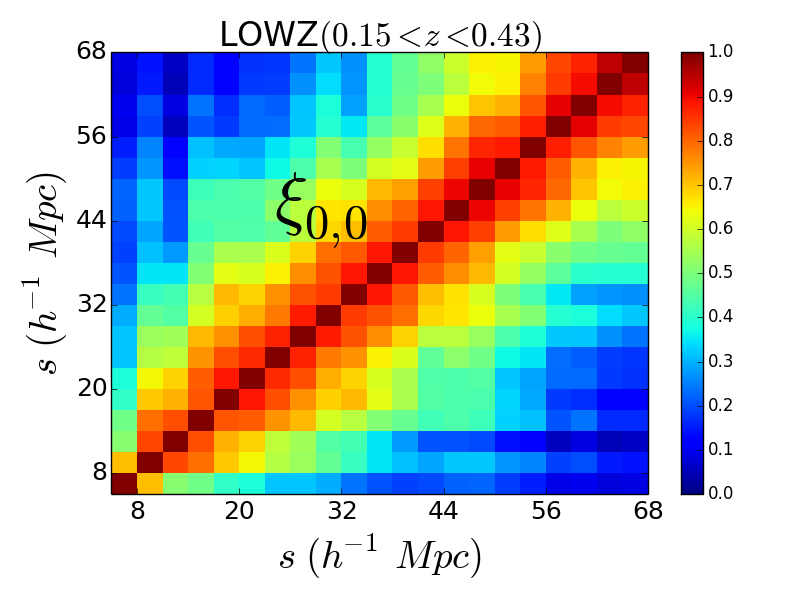}
    \includegraphics[width=0.45\textwidth]{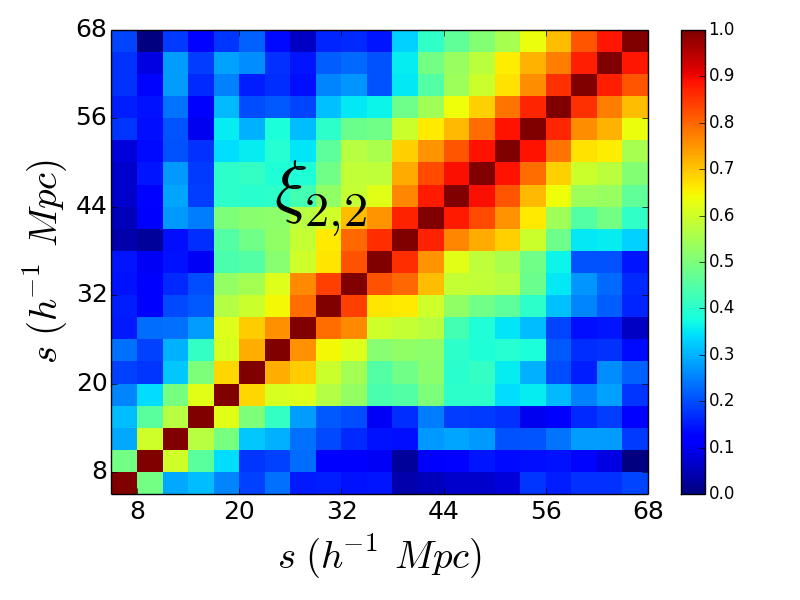}
    \caption{The plot on the top of the figure shows the correlation matrix for monopoles generated from 100 QPM mocks with $z_{\rm eff}=0.32$. The plot on the bottom of the figure shows the correlation matrix obtained for quadrupoles obtained from 100 QPM mocks with $z_{\rm eff}=0.32$. Both the correlation matrices have bins of sizes $3 \ h^{-1}$Mpc.}
    \label{fig:CorrXi}
\end{figure}

\subsection{Marked correlation functions} \label{sec:MarkCorrFunc}
Having discussed the design of the standard correlation function $\xi_{\ell}(s)$ and the mark weighted correlation function $W_{\ell}(s)$ in sections~\ref{sec:TwoPCorrFunc} and~\ref{sec:MarkWtCorrFunc} respectively, we now consider the structure of the `marked correlation function $\mathcal{M}_{\ell}(s)$'. If each galaxy in a given catalogue is assigned a mark $m$ based on the formula given in equation~\ref{eqn:mark}, then the marked correlation function $\mathcal{M}_{\ell}(s)$ can be defined by measure described in equation~\ref{eqn:MarkCorrFunc_Isotropic}. 
\begin{equation}
\label{eqn:MarkCorrFunc_Isotropic}
\mathcal{M}_{\ell}(s) \equiv \frac{1+W_{\ell}(s)}{1+\xi_{\ell}(s)}.
\end{equation} 

One prominent feature in the form of marked correlation function presented in equation~\ref{eqn:MarkCorrFunc_Isotropic} is that it ensures that the estimator effectively computes the correlation of the properties of galaxies with their respective environments \citep[][]{Sheth2005, Skibba2006, White2009, Skibba2009, Skibba2012}. Another significant attribute of marked correlation functions that is worth noticing is that, one would expect the estimator $\mathcal{M}_{\ell}(s)$ to approach the value 1 at large values of pair separation $s$. More details on the inspiration behind the choice of the definition of $\mathcal{M}_{\ell}(s)$ presented in equation~\ref{eqn:MarkCorrFunc_Isotropic} is given in \citet{Sheth2005} and \citet{White2016}. Figures~\ref{fig:LOWZ_Mark_Diff_p} and~\ref{fig:LOWZ_Mark_Diff_rhoratio} show plots of marked correlation functions which are obtained when marks with different values of parameters $p$ and $\rho_{ \star }$ are used to obtain mark weighted correlation functions.

\section{Analysis} \label{sec:Analysis}
In this section, we outline the techniques that we use for the statistical analysis of the correlation functions of BOSS DR12 LOWZ galaxies and galaxies from LOWZ QPM mock catalogues and subhalo catalogues from \textsc{elephant} GR simulation. 

\subsection{The covariance matrix} \label{sec:Covariance}
The analysis presented in this paper is based on comparison of standard and marked correlation function monopoles ($\xi_0(s)$ and $\mathcal{M}_{0}(s)$) from BOSS DR 12 LOWZ galaxy catalogue (observations) and catalogue from \textsc{elephant} GR simulation (theory). We compare data and theory using the $\chi^2$ statistic, using covariance matrices. In our analysis, we use LOWZ QPM mocks ($z_{\rm eff}=0.32$) to obtain estimates of covariance matrices for these standard and marked correlation function multipoles. For calculation of covariance matrices from the marked correlation function multipoles, we follow the prescriptions outlined in \citet{Vargas2013, Percival2014, Sidd2017}. 

\begin{align}
\hat{\Sigma}_{ij}^{\xi} =& \left[ \displaystyle\sum_{n=1}^{N_{\rm mock}}(\tilde{  \xi  }_{i,n} - \bar{ \xi })(\tilde{ \xi }_{j,n} - \bar{ \xi }) \right] \Bigg/ \left[ N_{\rm mock}-1 \right]. \label{eqn:CovarianceEqn_Xi} \\
\hat{\Sigma}_{ij}^{\mathcal{M}} =& \left[ \displaystyle\sum_{n=1}^{N_{\rm mock}}(\tilde{  \mathcal{M}  }_{i,n} - \bar{ \mathcal{M} })(\tilde{ \mathcal{M} }_{j,n} - \bar{ \mathcal{M} }) \right] \Bigg/ \left[ N_{\rm mock}-1 \right]. \label{eqn:CovarianceEqn_M}
\end{align}

In equations~\ref{eqn:CovarianceEqn_Xi} and~\ref{eqn:CovarianceEqn_M}, the indices `$i,j$' represent the indices of the binned values of the radial positions ($s$) in the standard and marked correlation function multipoles. Similarly, $\hat{\mathbf{\Sigma}}_{ij}^{\xi}$ and $\hat{\mathbf{\Sigma}}_{ij}^{\mathcal{M}}$ correspond to the $(i,j)$th entry of the computed covariance matrices for $\xi(s)$ and $\mathcal{M}(s)$ respectively. $N_{\rm mock}$ denotes the total number of mocks used in the analysis. We base all our investigations on 100 LOWZ QPM mocks. Thus, $N_{\rm mock}=100$ here. The symbol `$n$' refers to the index of the mock under consideration while $\bar{ \mathcal{M} }$ represents the mean of the marked correlation function multipoles obtained from the 100 mocks.

We can use the correlation matrix to express the relation between correlation functions obtained from the different mocks.  Equation~\ref{eqn:StatCorr} outlines the procedure for obtaining the correlation matrix $\hat{\mathbf{r}}$ from the covariance matrices $\hat{\mathbf{\Sigma}}^{\lbrace \xi, \mathcal{M} \rbrace}$.

\begin{equation}
\label{eqn:StatCorr}
\hat{\mathbf{r}}_{ij}^{\lbrace \xi, \mathcal{M} \rbrace} = \frac{\hat{\Sigma}_{ij}^{\lbrace \xi, \mathcal{M} \rbrace}}{ \sqrt{ \hat{\Sigma}_{ii}^{\lbrace \xi, \mathcal{M} \rbrace}\hat{\Sigma}_{jj}^{\lbrace \xi, \mathcal{M} \rbrace} }}.
\end{equation}

Fig.~\ref{fig:CorrXi}, shows plots of the monopole and quadrupole correlation matrices that are obtained from analysis of 100 mocks from the LOWZ redshift bin of the QPM mock catalogues.

\begin{figure}
    \includegraphics[width=0.41\textwidth]{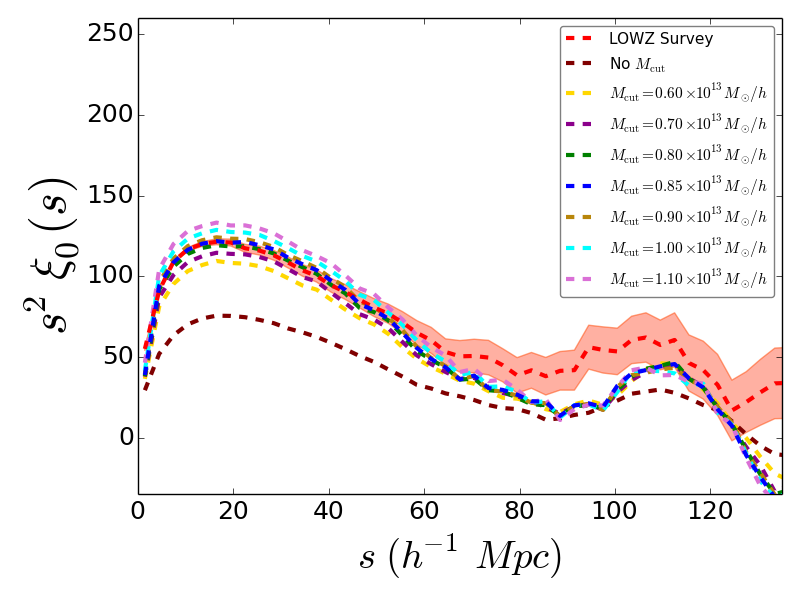}
    \includegraphics[width=0.41\textwidth]{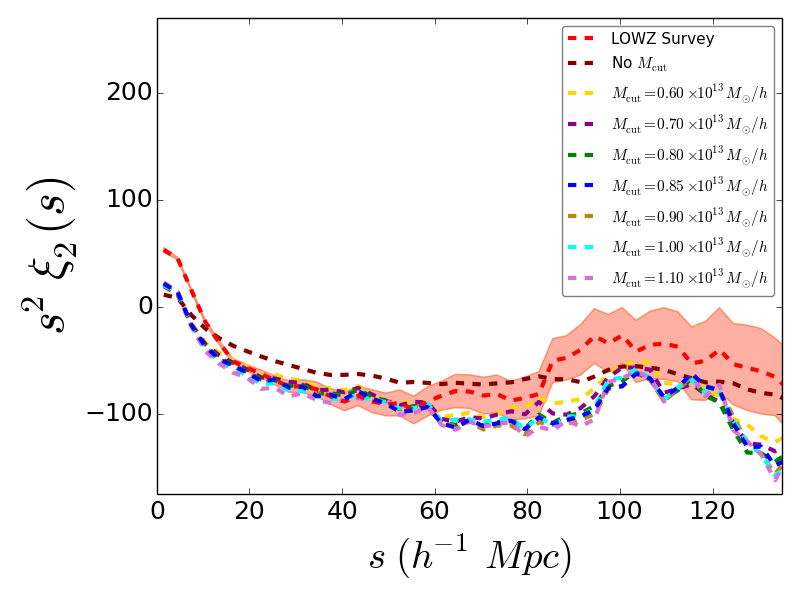} 
    \caption{We show the standard multipoles obtained from the subhalo catalogue of \textsc{elephant} GR simulation for various values of mass cuts. The top figure illustrates the monopoles for various values of mass cuts, while the bottom figure depicts plots of quadrupoles for different values of mass cuts. We also plot the (standard) monopole and quadrupole computed from the analysis of LOWZ galaxies in red dashed lines in the top and the bottom figures respectively. The red shaded regions denote the $1 \ \sigma$ errors obtained from the diagonal entries of the covariance matrices of 100 LOWZ QPM multipoles. From a $\chi^2$ analysis of the LOWZ monopole and monopoles corresponding to different values of mass cut in the subhalo catalogue of \textsc{elephant} GR simulation, we find that the monopole corresponding to mass cut of $0.85 \times 10^{13} M_{\odot}/h$ has the best fit with the LOWZ monopole. The effective mass of GR subhaloes with mass cut $M_{\rm cut} = 0.85 \times 10^{13} M_{\odot}/h$ is $M_{\rm eff} = 2.75 \times 10^{13} M_{\odot}/h$. }
    \label{fig:RockstarMassCut}   
\end{figure}

\begin{figure}
    \includegraphics[width=0.41\textwidth]{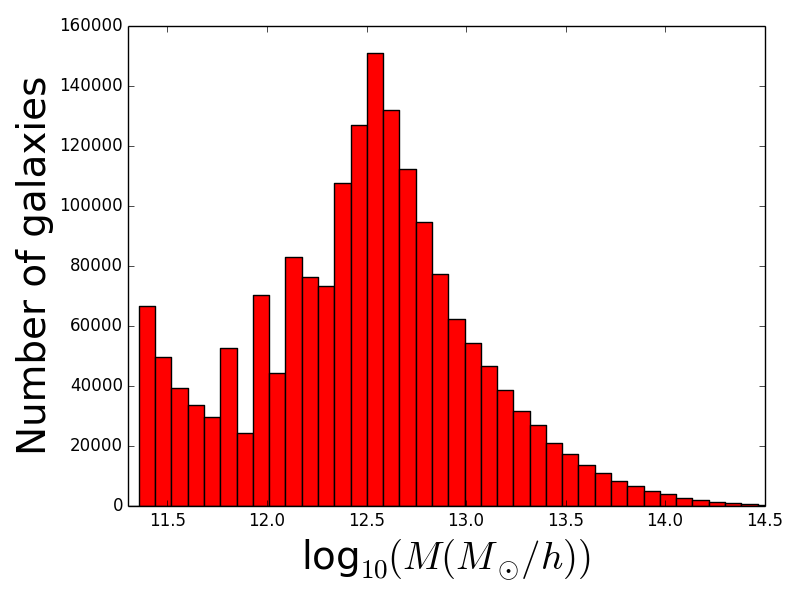}
    \caption{The histogram shown here depicts the distribution of masses of subhaloes in the simulation GR catalogue ($z_{\rm eff}=0.3470$).}    
    \label{fig:Rockstar_MassDist}
\end{figure}

\subsection{Model comparison} \label{sec:Chi2}
Using both the standard and marked correlation function estimators we check the agreement of observations (BOSS DR 12 LOWZ galaxies) and theory (GR simulation subhalo catalogue), using the $\chi^2$ statistic. In our analysis, we assume the errors on the standard correlation function multipoles to be Gaussian distributed. For estimators of any kind (standard correlation functions or marked correlation functions), we first construct the data vectors ($\xi^{\rm data}$ or $\mathcal{M}^{\rm data}$) from the monopoles of the given estimator ($\xi_0$ or $\mathcal{M}_0$). These data monopoles are obtained from analysis of BOSS DR12 LOWZ galaxies. We repeat the same prescription for the standard and marked correlation function monopoles obtained from GR simulation subhalo catalogue to get the theory vector ($\xi^{\rm theory}$ or $\mathcal{M}^{\rm theory}$). That is, $\xi^{\rm data} = \tilde{\xi}^{\rm data}_{0}$, $\mathcal{M}^{\rm data} = \tilde{\mathcal{M}}^{\rm data}_{0}$, $\xi^{\rm theory} = \tilde{\xi}^{\rm theory}_{0}$ and $\mathcal{M}^{\rm theory} = \tilde{\mathcal{M}}^{\rm theory}_{0}$. 

Once we have the data vector $\xi^{\rm data}$ (or, $\mathcal{M}^{\rm data}$), the theory vector $\xi^{\rm theory}$ (or, $\mathcal{M}^{\rm theory}$) and the covariance matrix $\mathbf{\Sigma}^{\xi}$ (or, $\mathbf{\Sigma}^{\mathcal{M}}$), we obtain the relevant $\chi^2$ statistic from equation~\ref{eqn:chi2}.

\begin{align}
\label{eqn:chi2}
\chi^2_{\xi} =& \left( \xi^{\rm data} - \xi^{\rm theory} \right) \left( \Sigma^{\xi} \right)^{-1} \left( \xi^{\rm data} - \xi^{\rm theory} \right)^T.  \\
\chi^2_{\mathcal{M}} =& \left( \mathcal{M}^{\rm data} - \mathcal{M}^{\rm theory} \right) \left( \Sigma^{\mathcal{M}} \right)^{-1}  \left( \mathcal{M}^{\rm data} - \mathcal{M}^{\rm theory} \right)^T.
\end{align}

A complete analysis of the goodness of fit between the data and the theory vectors requires the use of reduced $\chi^2$ statistic, $\chi^2/dof$, where $dof$ is the degrees of freedom. The $dof$  can be computed from knowledge of the bin size, minimum ($s_{\rm min}$) and maximum ($s_{\rm max}$) scales of fitting. We use evenly spaced bins of $s$ with bin sizes of $s=3 \ h^{-1}$Mpc in our analysis. For the minimum ($s_{\rm min}$) and maximum ($s_{\rm max}$)  fitting scales in the $\chi^2$ analysis, we use values of $6 \ h^{-1}$Mpc and $69 \ h^{-1}$Mpc respectively, and the relation $dof = \left( s_{\rm max} - s_{\rm min} \right) / {\rm bin \ size}$. In our work, we choose theory vectors after $\chi^2$ analysis of standard correlation function monopoles obtained from various mass cuts of GR simulation subhalo catalogue. Fig.~\ref{fig:RockstarMassCut} shows multipoles obtained from various mass cuts of GR simulation. We discuss the choice of the prefered (`effective') mass cut and theory correlation multipoles in detail in section~\ref{sec:Results}.

\begin{figure}
    \includegraphics[width=0.41\textwidth]{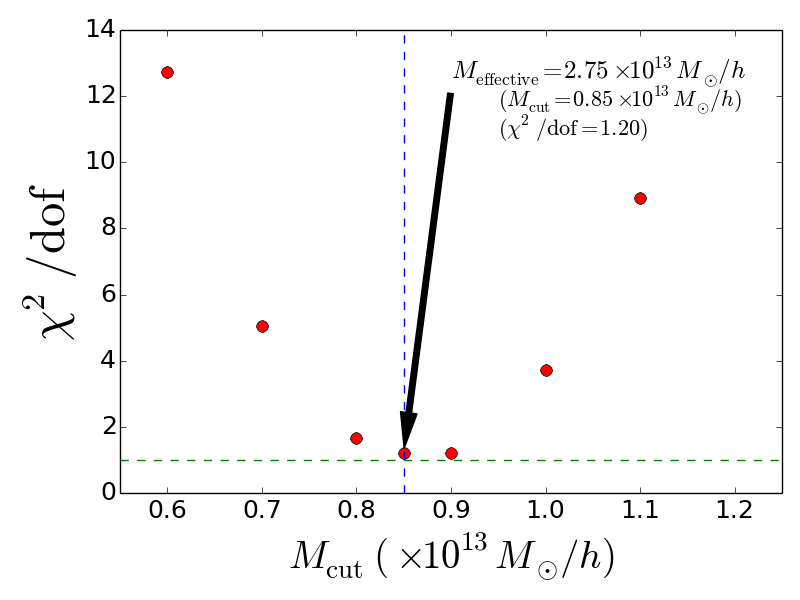}
    \caption{This figure plots values of $\chi^2/{\rm dof}$ obtained in the comparison of the LOWZ monopole with different monopoles computed with the consideration of various mass cuts in the catalogue of GR simulation subhaloes. We find that the best fit value of mass cut corresponds to $M_{\rm cut} = 0.85 \times 10^{13} M_{\odot}/h$. This is corroborated by the plots of monopoles shown in Fig.~\ref{fig:RockstarMassCut}}.
    \label{fig:RockstarMassCut_Chi2}     
\end{figure}

\begin{figure}
    \includegraphics[width=0.41\textwidth]{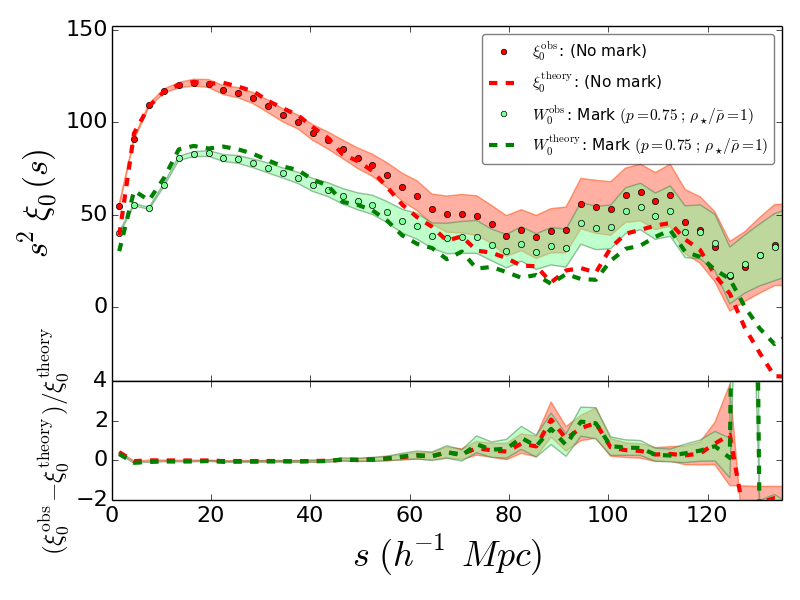}
    \includegraphics[width=0.41\textwidth]{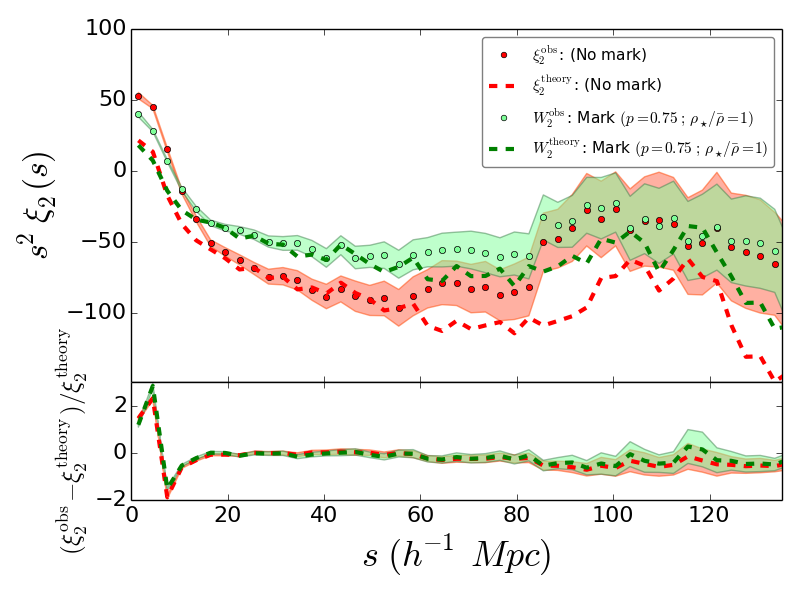}
    \caption{Plots of standard isotropic galaxy correlation functions obtained from the LOWZ galaxy observations. These standard galaxy correlation functions are computed using the weights discussed in equation~\ref{eqn:Anderson} and are depicted as red dots (observation multipoles, $\xi_{0}^{\rm obs}$, $\xi_{2}^{\rm obs}$). Plots in the top figure depict monopoles, whereas all relevant quadrupoles are shown in the bottom figure. The red shaded regions around the observation multipoles denote the $1 \ \sigma$ error obtained from analysis of 100 QPM mocks ($z_{\rm eff}=0.32$) based on the use of the weighting scheme described equation~\ref{eqn:Anderson}. In both the top and bottom figures, we have shown mark weighted correlation functions with a specific choice of parameters, $viz.$ $p=0.75, \rho_{ \star } / \bar{\rho}=1$. These mark weighted correlation functions are shown in the form of green colored dots. The shaded green regions in both figures denote the $1 \ \sigma$ errors gotten from the survey of mark weighted correlation function multipoles obtained from 100 LOWZ QPM mocks. In both the figures, the red dashed lines represent the best fit standard multipoles ($\xi_{0}^{\rm theory}$, $\xi_{2}^{\rm theory}$) obtained from GR simulation subhalo catalogue. Mark weighted correlation function multipoles from the GR simulation subhalo catalogue which correspond to the best fit standard multipoles ($\xi_{0}^{\rm theory}$, $\xi_{2}^{\rm theory}$) are denoted as dashed green lines ($W_{0}^{\rm theory}$, $W_{2}^{\rm theory}$).}    
    \label{fig:LOWZ_Corr_ChosenParam}
\end{figure} 

\begin{figure}
    \includegraphics[width=0.41\textwidth]{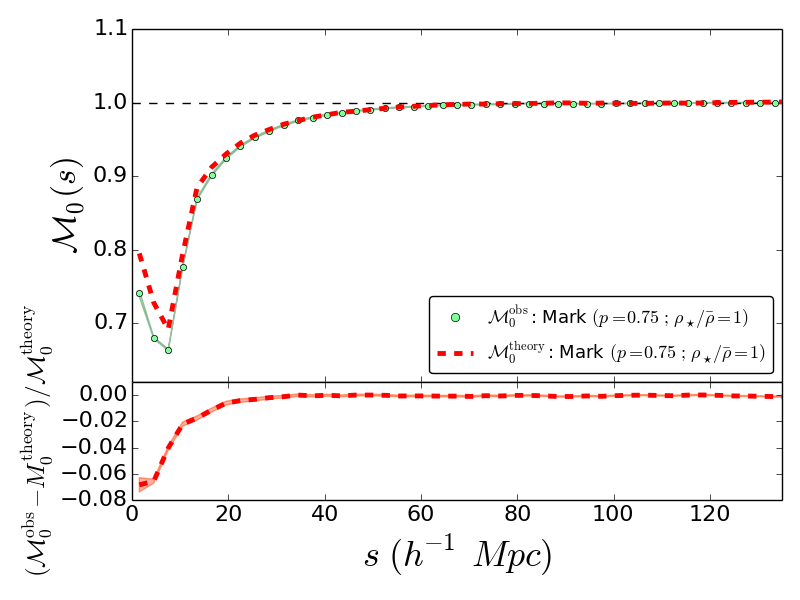}
    \includegraphics[width=0.41\textwidth]{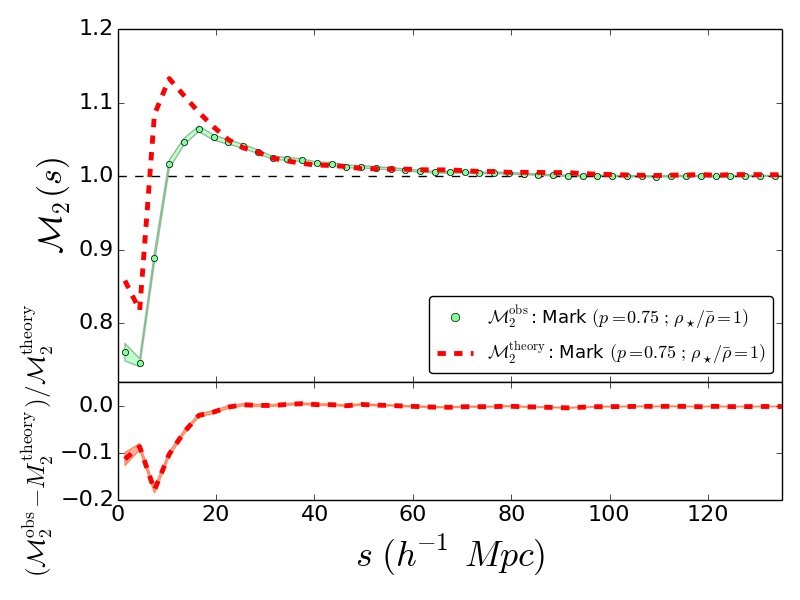}
    \caption{We illustrate the marked correlation function multipoles of LOWZ galaxies for the chosen value of mark $m$ ($p=0.75, \rho_{ \star } / \bar{\rho}=1$) in these figures. The top and bottom figures illustrate the marked correlation function monopole and quadrupole respectively. The green dots in both the plots represent marked correlation function multipoles for the LOWZ DR12 galaxy sample. The green shaded regions in the plots denote the $1\sigma$ errors obtained from the diagonal entries of the covariance matrices which are gotten from the analysis of 100 QPM mocks. The red dashed lines in the top and the bottom figures represent the theory marked correlation functions obtained from the best fit correlation function multipoles from GR simulation subhalo catalogue.}
    \label{fig:LOWZ_Mark_ChosenParam}
\end{figure} 

\begin{figure}
    \includegraphics[width=0.41\textwidth]{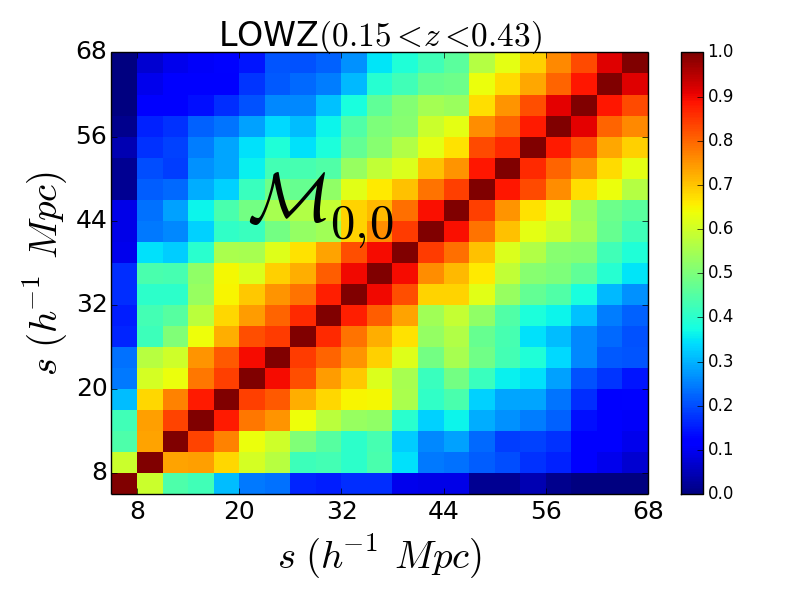}
    \includegraphics[width=0.41\textwidth]{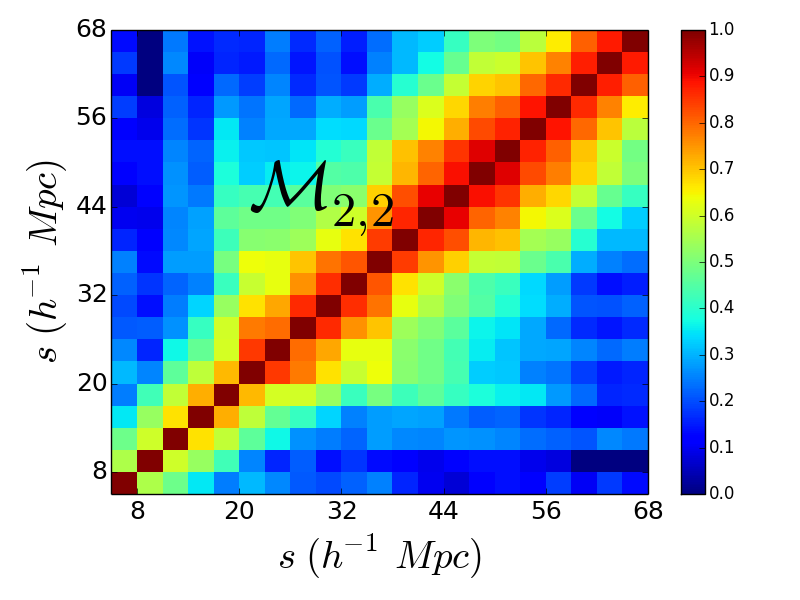}
    \caption{These plots show correlation matrices of marked correlation function monopoles and quadrupoles, which are gotten from the analysis of 100 LOWZ QPM mocks. The bins shown in these correlation matrices have sizes of $3 \ h^{-1}$Mpc.}
    \label{fig:LOWZ_StatCorr}
\end{figure}

\section{Results} \label{sec:Results}
We discuss the results that we have obtained for standard, mark weighted and marked correlation functions from BOSS DR12 galaxy data, GR simulation subhalo catalogue and LOWZ QPM mocks in this section. We also give details of model comparison between estimators (standard and marked correlation functions) from theory and observation.
\subsection{Theory correlation functions from GR simulation} \label{sec:TheoryGRRockstar}

\begin{table}
	\centering
	\caption{The first column gives the mass cut of subhaloes in the GR simulation subhalo catalogue (in units of $ 10^{13} h^{-1} M_{\odot}$). The second column gives the effective mass of subhaloes (in units of $ 10^{13} h^{-1} M_{\odot}$) with mass higher than the corresponding mass cut. We show the correlation function multipoles obtained from the various mass cut samples in Fig.~\ref{fig:RockstarMassCut}.}
	\label{tab:Eff_Mass_Rockstar}
	\begin{tabular}{ccr}
		\hline
		\hline
		$M_{\rm cut} \ (  h^{-1} M_{\odot} )$& $M_{\rm eff} \ ( h^{-1} M_{\odot} )$ & Number of subhaloes \\
		\hline
		$0 $ & $ 7.18 \times 10^{12}$ & 1,902,278  \\			
		$0.60 \times 10^{13}$ & $ 2.12 \times 10^{13} $ & 495,580 \\	
		$0.70 \times 10^{13}$ & $ 2.38 \times 10^{13} $  & 421,462 \\	
		$0.80 \times 10^{13}$ & $ 2.63 \times 10^{13} $  & 366,348 \\	
		$0.85 \times 10^{13}$ & $ 2.75 \times 10^{13} $  & 341,941 \\	
		$0.90 \times 10^{13}$ & $ 2.86 \times 10^{13} $  & 323,597 \\
		$1.00 \times 10^{13}$ & $ 3.09 \times 10^{13} $ & 289,133 \\					
		$1.10 \times 10^{13}$ & $ 3.31 \times 10^{13} $ & 260,894  \\										
		\hline
	\end{tabular}
\end{table}

If the mean of correlation functions obtained from all QPM mocks is taken as the theory, then one finds that there is not a very good match between theory and observation correlation functions. In particular, if the mean $\xi_0$ obtained from all QPM mocks is taken as the theory, then we find that a comparison between $\xi_0^{\rm obs}$ and $\xi_{\rm 0, \ mean \ of \ mocks}^{\rm theory}$ gives $\chi^2/dof \sim 11$. This is the reason why we opt to have theory multipoles from the GR simulation catalogue.

The distribution of masses of subhaloes in the GR simulation catalogue is shown in the form of a histogram in Fig.~\ref{fig:Rockstar_MassDist}. The effective mass of the subhaloes in the GR simulation catalogue is $7.18 \times 10^{12} M_{\odot}/h$. This is lower than the effective mass of BOSS DR 12 LOWZ galaxies. Hence, while obtaining correlation function multipoles in the GR simulation catalogue, we use mass cutoffs ($M_{\rm cut}$) and select subhaloes above the specified mass cutoffs to obtain the correlation function multipoles. This exercise serves to increase the amplitude of the multipoles from the GR simulation subhalo catalogue. Results of multipoles obtained from different mass cuts of GR simlulation are depicted in Fig.~\ref{fig:RockstarMassCut}. To obtain the preferred value of mass cut, we compare monopoles obtained from mass cuts of GR simulation with monopole from BOSS DR12 LOWZ galaxies (in the chosen distance scale $s_{\rm min} = 6 \ h^{-1}$Mpc and $s_{\rm max} = 69 \ h^{-1}$Mpc) using $\chi^2$ analysis discussed in section~\ref{sec:Chi2}. Fig.~\ref{fig:RockstarMassCut_Chi2} shows values of $\chi^2_{\xi_0}/dof$ that are obtained from the comparison of monopoles from different mass cuts of GR simulation with monopole from BOSS DR 12 LOWZ galaxies in the chosen distance scale. We select the mass cut corresponding to the GR monopoles which have the best matching with monopoles from LOWZ galaxies as the preferred mass cut. From our analysis, we find that the GR monopole with the best goodness of fit with BOSS DR12 LOWZ monopole (lowest $\chi^2_{\xi_0}/dof$) between the scales $s_{\rm min} = 6 \ h^{-1}$Mpc and $s_{\rm max} = 69 \ h^{-1}$Mpc corresponds to $M_{\rm cut} =  0.85 \times 10^{13} M_{\odot}/h$. The comparison between the GR monopole from a sample corresponding to mass cut of $M_{\rm cut} =  0.85 \times 10^{13} M_{\odot}/h$ and LOWZ monopole in the said fitting range, yields a value of $\chi^2_{\xi_0}/dof=1.20$. This mass cut of $M_{\rm cut} =  0.85 \times 10^{13} M_{\odot}/h$ corresponds to an effective mass of $M_{\rm eff} =  2.75 \times 10^{13} M_{\odot}/h$. In Fig.~\ref{fig:LOWZ_Corr_ChosenParam}, we show plots of standard correlation function multipoles from BOSS DR12 LOWZ galaxy catalogue and GR simulation subhaloes chosen from a sample corresponding to the effective mass cut. We find marked correlation functions from the GR subhalo catalogue using the effective mass cut and present analysis of the same in section~\ref{sec:MarkedCorrrResult}.

\subsection{Marked correlation functions} \label{sec:MarkedCorrrResult}
Having gotten the best fit mass cut for the \textsc{elephant} GR simulation subhalo catalogue ($M_{\rm cut} =  0.85 \times 10^{13} M_{\odot}/h$), we are now in a position to perform a $\chi^2/dof$ goodness of fit between the marked correlation function monopole obtained from BOSS DR12 LOWZ galaxies and the marked correlation function monopole obtained for the best fit mass cut of GR subhalo catalogue between the distance ranges $s_{\rm min} = 6 \ h^{-1}$Mpc and $s_{\rm max} = 69 \ h^{-1}$Mpc. Fig.~\ref{fig:LOWZ_Mark_ChosenParam} shows marked correlation function multipoles obtained from LOWZ galaxies and GR simulations. In Fig.~\ref{fig:LOWZ_StatCorr}, we illustrate covariance matrices of marked correlation function multipoles obtained from 100 QPM mocks at $z_{\rm eff}=0.32$. 

The comparison of marked correlation monopoles obtained from LOWZ galaxies with those obtained from the best fit mass cut of GR subhaloes yields a $\chi^2_{\mathcal{M}_0}/dof=40.68$. Strong correlations between standard and mark weighted correlated functions are one of the major reasons for this high value of $\chi^2/dof$. This, and other reasons which can be attributed to features in the design of the marked correlation function are discussed in detail in appendix~\ref{sec:chi2markedcorr}. 

One possibility for the high value of  $\chi^2_{\mathcal{M}_0}/dof$ might be the use of low resolution QPM mocks in the estimation of the covariance matrix $\Sigma_{\mathcal{M}_0}^{\rm QPM}$. We estimate the covariance matrix $\Sigma_{\mathcal{M}_0}$ using jackknife regions obtained from \textsc{Elephant} GR simulation catalogs to check if the mentioned possibility is true. In particular, we obtain 125 different jackknife regions by masking regions of size 204.8 Mpc/h (in x,y and z axes) in the GR simulation catalogue (total original volume $1024\times1024\times1024$ Mpc$^3$/h$^3$). From these 125 regions, we obtain 125 separate standard multipoles ($\xi_{0,2}^{\rm Jackknife}$), mark weighted multipoles ($W_{0,2}^{\rm Jackknife}$) and marked correlation function multipoles ($\mathcal{M}_{0,2}^{\rm Jackknife}$). Using the 125 separate marked correlation function monopoles ($\mathcal{M}_{0}^{\rm Jackknife}$), we form another covariance matrix ($\Sigma_{ \mathcal{M}_0 }^{\rm Jackknife}$) and use that for estimation of $\chi^2_{\mathcal{M}_0}/dof$. For the correlation function obtained from jackknife regions, we again get a high value of $\chi^2_{\mathcal{M}_0}/dof = 2149$. Here also, we find that $W$ and $\xi$ vectors are very highly correlated. The jackknife analysis is described in detail in appendix~\ref{sec:chi2markedcorr_jn}.

It is likely that $\mathcal{M}_0$ being a ratio is not sensitive to cosmic variance (much like \citet{McDonald2009} ) . This could in principle be a good thing; but it also means that the theory has to be very accurate. In this case, the weak link is assigning galaxies to the mass in the simulation. One would need to do better here. This is effectively a systematic error that doesn't show up in the covariance matrices that we have used.  

It is still possible to make a robust comparison between observation and theory multipoles using an analysis based on fractional deviation (bottom panels of Fig.~\ref{fig:LOWZ_Mark_ChosenParam}). We discuss this is detail in section~\ref{sec:Implications}.

\section{Discussion and Conclusions} \label{sec:Discussion}
\subsection{Implications for tests of f(R) gravity}  \label{sec:Implications}
We compare our work to the recent investigations of $f$(R) models reported in \citet{Armijo2018, Hernandez-Aguayo2018}. There are some significant differences between the methodologies adopted. First, we have computed two point and marked correlation functions in redshift space, while these cited investigations are based on regular two point and marked correlation functions computed in real space. In \citet{Armijo2018, Hernandez-Aguayo2018}, the matching of correlation functions is in real space with no redshift space distortions \citep[][]{Cautun2017}. We do not compute correlation functions in real space in our paper. Also, in \citet{Armijo2018, Hernandez-Aguayo2018}, the authors uncover differences in the correlation functions on smaller scales than the scales that we have probed. Another difference between our works is that the previous authors use several approaches to compute marks for the marked correlation function estimators. Among the different methods that they use to compute the effects of local environments around galaxies, marks based on subhalo mass, local density and gravitational potential feature prominently. Our procedure for computing marks (described in section~\ref{sec:MarkWtCorrFunc}) is similar in spirit, but not in detail as the techniques that have been used in \citet{Armijo2018, Hernandez-Aguayo2018}. Also, the galaxy data for which we compute correlation functions are at redshift $z \sim 0.3$, while the work in \cite{Hernandez-Aguayo2018} focuses on halos and HOD galaxies at $z=0.5$.

For LOWZ observations and the GR simulations, we find that the mean variation between the standard monopoles ($\xi_0$) in the relevant distance scale ($6 \ h^{-1}$Mpc $\leq  s  \leq$ $69 \ h^{-1}$Mpc) is $ \sim 5.3 \%$, the minimum variation is $\sim 0.10 \%$ and the maximum variation is $\sim 23 \%$. In our investigations, we find that it is difficult to find a good match between the quadrupoles of LOWZ observations and GR simulation in redshift space. The best match obtained between the quarupoles ($\xi_2$) over the relevant distance scale ($6 \ h^{-1}$Mpc $\leq  s  \leq$ $69 \ h^{-1}$Mpc) is $\sim 0.20 \%$, the mean variation is $\sim 19 \%$ and the maximum variation is $\sim 188 \%$. 

At the same time, we find that the mean deviation between the marked correlation function monopoles ($\mathcal{M}_0$) obtained for LOWZ observations and GR simulation in the aforementioned distance scale is $0.55 \%$, the minimum deviation is $0.00065 \%$ and the maximum deviation is $ 4.0 \%$. The deviations of marked correlation function quadrupoles ($\mathcal{M}_2$) are higher when compared to the deviations of marked correlation function monopoles. In the said distance range, the mean deviation between the marked correlation function quadrupoles is found to be is $ 2.0 \%$, the minimum deviation is found to be $0.0052 \%$, while the maximum deviation is found to be is $17 \%$. We find that the highest deviations occur at the smallest distance scale ($\sim 6 \ h^{-1}$Mpc) considered by us. Given these results for the marked correlation function multipoles, we find no evidence of deviation with respect to $\Lambda$CDM+GR. 

\citet{Armijo2018, Hernandez-Aguayo2018} report results of deviations between marked correlation functions of several $f$(R) gravity models and GR simulations. These authors find, as we do, that deviations between marked correlation functions of modified gravity models and $\Lambda$CDM+GR are largest on scales $< 10 \ h^{-1}$Mpc. Their work presents results for several different models and many choices of mark. On the smallest scale we measure here ($6 h^{-1}$Mpc), these authors find a maximum difference of $\sim10\%$ for the model with the greatest difference to $\Lambda$CDM. The majority of models exhibit differences of a few percent or less. As there are differences between our works (mentioned above) we do not do a direct comparison. 

In this type of analysis, the unmarked correlation functions of simulated subhaloes must be matched (in redshift space) with those of the observations before comparing the marked correlation functions. This means that this matching of unmarked correlation functions limits the precision of these tests
of modified gravity. The mean error in matching (quoted above) is $5.3\%$, which is larger than the expected effects of most modfied gravity models on the marked correlation function. We are therefore not able to rule out  modified gravity models yet. The statistical errors on the observational measurements are however $5.2\%$ on scales of $6 \ h^{-1}$Mpc $\leq  s  \leq$ $69 \ h^{-1}$Mpc, meaning that such tests will be possible with improvements in matching the unmarked correlation functions.

\subsection{Conclusions}
We have investigated the agreement of General Relativity with BOSS DR 12 observations with the use of marked correlation functions proposed in \citet{White2016}. We have shown that the consideration of local densities around galaxies and the use of marked correlation functions leads to no evidence of deviation between $\Lambda$CDM universe and BOSS DR 12 LOWZ observations on distance scales $6 \ h^{-1}$Mpc $ \leq s \leq$ $69 \ h^{-1}$Mpc. There are significant challenges in RSD analysis of small distance scales. In our choice of distance scale for model comparison, we include quasi-linear and non-linear scales, where the presence of peculiar velocities becomes very important. Virialized cluster velocities \citep[][]{Jackson1972, Tegmark2004} could be playing a significant role in our comparison of  results.

Although our work reveals no evidence of deviations between GR models and observations, we realize the need for better modeling in redshift space. Further work has to be done in this regard. Consideration of halo models and better modeling in redshift space are likely to give more robust answers. At the same time, we have made a measurement in this work which can be compared to by research groups which are working on similar marked correlation function estimators with comparable or better modeling in redshift space. 

If the accurate modelling of redshift distortions on small scales proves to be a limit on the constraining power of marked correlation functions, one could imagine working with projected clustering or angular clustering. There would still be issues to overcome, however, as for the mark to have physical relevance (for example being related to an actual density) it would likely need to be computed in three spatial dimensions rather than from the angular or projected density. On the other hand, the latter may still be good enough to reveal significant differences between models without dealing with redshift distortions, and this should be investigated.

Our work has demonstrated a technique for the computation of marked correlation functions. The method that we have outlined for the computation of marked correlation functions necessitates very little change to presently exisiting pipelines for the computation of clustering statistics. As such, our method and results can be easily replicated and employed in similar works in the future. Also, we hope that our observational measurements will be useful to constrain and guide similar works in the future.

\section*{Acknowledgements}
The authors acknowledge helpful comments and suggestions from Antonio J. Cuesta, Alex Geringer-Sameth, Manfred Paulini, Jeremy Tinker, Mariana Vargas-Maga\~{n}a and Hongyu Zhu. 

RACC is supported by a Lyle Fellowship from the University of Melbourne. 

SS thanks Aileen Zhai and Oshadha Gunasekara for useful suggestions that greatly contributed to the research. 

BL acknowledges support by the European Research Council (ERC) through grant ERC-StG-716532-PUNCA and the UK Science and Technology Facilities Council (STFC) through grant ST/P000541/1. Funding for SDSS-III has been provided by the Alfred P. Sloan Foundation, the Participating Institutions, the National Science Foundation, and the U.S. Department of Energy Office of Science. The SDSS-III web site is \href{http://www.sdss3.org/}{http://www.sdss3.org/}. 

SDSS-III is managed by the Astrophysical Research Consortium for the Participating Institutions of the SDSS-III Collaboration including the University of Arizona, the Brazilian Participation Group, Brookhaven National Laboratory, Carnegie Mellon University, University of Florida, the French Participation Group, the German Participation Group, Harvard University, the Instituto de Astrofisica de Canarias, the Michigan State/Notre Dame/JINA Participation Group, Johns Hopkins University, Lawrence Berkeley National Laboratory, Max Planck Institute for Astrophysics, Max Planck Institute for Extraterrestrial Physics, New Mexico State University, New York University, Ohio State University, Pennsylvania State University, University of Portsmouth, Princeton University, the Spanish Participation Group, University of Tokyo, University of Utah, Vanderbilt University, University of Virginia, University of Washington, and Yale University.

\bibliographystyle{mnras}
\bibliography{ref} 




\appendix

\section{Accounting for Observational Artefacts in BOSS galaxies} \label{sec:Systematics}

Below we discuss how galaxies in BOSS are weighted when computing large scale structure (LSS) statistics. The consideration of the weights is done in order to reduce the effect of observational artefacts on the estimate of the true galaxy overdensity field. There are various effects which need to be examined while assessing the galaxies spectroscopically observed in BOSS. Thorough scrutinies of these effects are presented in \citet{Anderson2012, Anderson2014} and \citet{Reid2016}. 

\subsection{Effect of fibre collisions}
In BOSS DR12 catalogue, there are galaxies which exist within the fiber collision radius ($62''$) of other targets. These galaxies are not allocated spectroscopic fibers due to ``fiber collisons". The galaxies which experience fiber collisions reside in denser environments, and consequently, they have higher than average large-scale
clustering. Such galaxies are also more likely to inhabit the same dark matter haloes as their neighbouring galaxy targets. These systematics and the requirement for accurate estimates of galaxy clustering drive the need to make corrections for fiber collisiions.

The method that is followed in default large scale structure catalogues focuses on upweighting the nearest galaxy in the same target class that was alloted a fiber with `fiber collided' galaxies. This upweighting accounts for neighbouring (collided) galaxies whose redshifts were not obtained since they were in close pairs . In BOSS DR12 catalogues, this weighting scheme is represented by the symbol $w_{\rm cp}$ (or, ${\rm WEIGHT}_{\rm CP}$). It is worth noting here, that these neighbours are upweighted without reference to their classification (good galaxy redshift, star or redshift), since the missed entity can belong to any of these categories. The ensuing fiber collision weights are extremely important on small scales while their effect on clustering is insignificant at the BAO scale.

\subsection{Effect of redshift failures}
Another cause for corrections in large scale structure statistics of galaxies arises when the spectroscopic pipeline is unsuccessful in acquiring redshifts of galaxy targets. Such targets are not expected to be distributed randomly with respect to redshifts or plate centers. Hence, we assume a nearest neighbour upweighting scheme while dealing with these galaxies. When the nearest neighbour of such a galaxy (of the same target class) has such a `redshift failure', the total weight is transferred to the nearest neighbour of the redshift failure. This correction of the effects of redshift failures leads to a new weight, $viz.$ $w_{\rm noz}$ (or, ${\rm WEIGHT}_{\rm NOZ}$) for galaxies in BOSS DR12. Approximately $0.5 \ \%$ of BOSS DR12 galaxies in the redshift bin $0.15 < z < 0.43$ and $1.8 \ \%$ of BOSS DR12 galaxies in the redshift bin $0.43 < z < 0.70$ benefit from this weighting scheme.

\subsection{Angular Systematic Weights}
A robust and complete analysis of large scale structure galaxy catalogues necessitates the consideration of corrective weights for stellar density ($w_{\rm star}$) and for seeing estimates ($w_{\rm see}$). A detailed description of the treatment of non-cosmological fluctuations in stellar density and seeing in BOSS galaxies is given in \citet{Ross2013}. 

The aforementioned non-cosmological fluctuations are characteristic of massive galaxies present in the CMASS sample. Since the LOWZ galaxies are brighter than the CMASS galaxies and because they do not exhibit significant variations in stellar density and seeing, they do not need these corrections. In other words, the values of $w_{\rm star}$ and $w_{\rm see}$ are equal to 1 for LOWZ galaxies. More details of the corrective weights $w_{\rm star}$ and $w_{\rm see}$ can be found in \citet{Ross2013}, \citet{Anderson2014} and \citet{Reid2016}. 

From the weights $w_{\rm star}$ and $w_{\rm see}$, one can obtain the total angular systematic weight $w_{\rm systot}$ as: $w_{\rm systot} = w_{\rm star} \times w_{\rm see}.$


\subsection{FKP Weights}
In 1994, Feldman, Kaiser and Peacock (hereafter abbreviated as FKP) suggested a weighting scheme \citep[][]{FKP1994} where galaxies are weighted based on the number density of galaxy tracers. We refer to this weight as $w_{\rm FKP}$ (or, ${\rm WEIGHT}_{\rm FKP}$). Since the number density of galaxy tracers depends on redshift, the weight assigned to each galaxy in this weighting scheme is a function of redshift. Also, this weight depends reciprocally on the amplitude of the power spectrum in the power spectrum bin of interest. Equation~\ref{eqn:FKP} outlines the method for obtaining the optimal weight $w_{\rm FKP}$ from the number density of galaxy tracers $\bar{n}(z)$ and the power spectrum $P_0$.

\begin{equation}
\label{eqn:FKP}
w_{\rm FKP, i} = \frac{1}{1+\bar{n}(z_i)P_0}.
\end{equation}

In BOSS DR12 galaxy catalogues, a value of $P_0=10000 \ h^{-3}$Mpc$^3$ is used to determine the power spectrum and evaluate the optimal $w_{\rm FKP}$ weights for galaxies. This chosen value of $P_0$ corresponds to the observed power spectrum at $k \simeq 0.15 \ h$ Mpc$^{-1}$. 

\citet{Percival2004} presents an updated version of the FKP weighting scheme which also considers luminosity dependent clustering. Because of the high efficiency of selection of massive galaxies in the BOSS DR12 catalogue, the gain provided by luminosity dependent weights is significant for galaxies in the BOSS DR12 catalogue.

  The use of FKP weights is optional but we have found the FKP weights useful for suppression of statistical errors in conjunction with the fiber collision, redshift failure and angular systematic weights as per the prescription suggested in \citet{Reid2016} and outlined in section~\ref{sec:AndWeight}.

\subsection{Total Galaxy Weights} \label{sec:AndWeight}
Following the considerations related to the fiber collision, redshift failure, angular systematic and FKP weights, it is natural to discuss a weighting scheme which combines all the aforementioned weights. In our work, we follow the combined weighting scheme proposed in \citet{Reid2016}. In it each galaxy is counted using a weight given in equation~\ref{eqn:Anderson}.
\begin{equation}
\label{eqn:Anderson}
w_{\rm tot} = \left( w_{\rm cp} + w_{\rm noz} - 1  \right) w_{\rm star} w_{\rm see} w_{\rm FKP}.
\end{equation}

The first part of equation~\ref{eqn:Anderson}, \textit{i.e.} the term in the parenthesis, ensures conservation
of the total number of galaxies. This weighting scheme gives unbiased estimates of the galaxy density field.

\section{$\chi^2$ analysis in marked correlation functions using QPM mocks} \label{sec:chi2markedcorr}

\begin{figure}
    \includegraphics[width=0.41\textwidth]{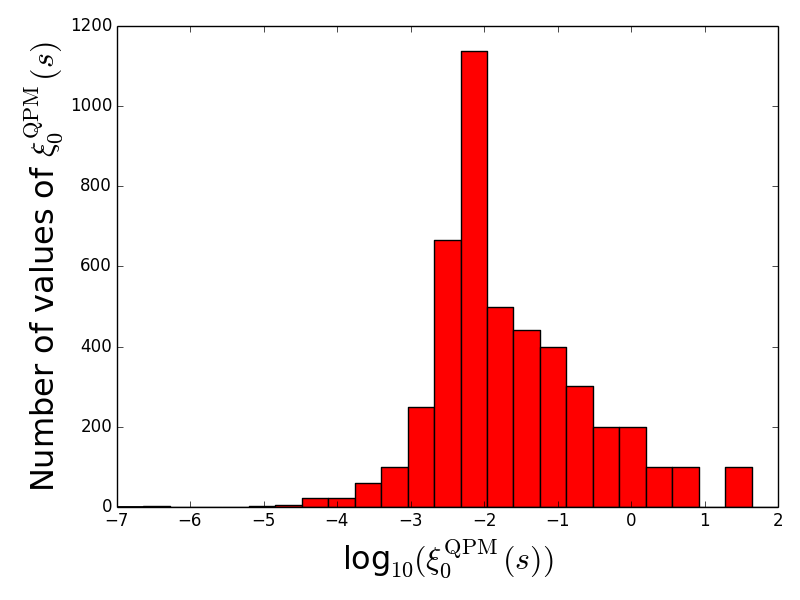}
    \includegraphics[width=0.41\textwidth]{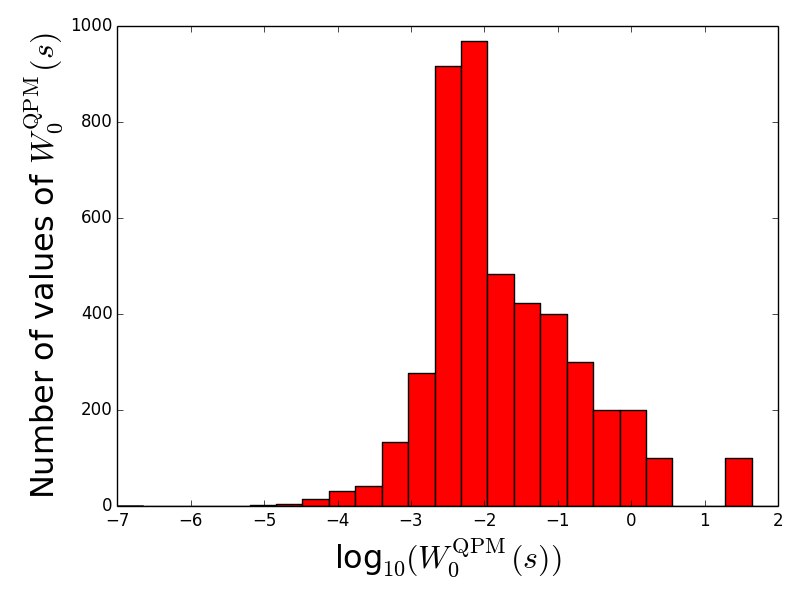}
    \caption{Here we plot distributions of values in 100 $\xi_0^{\rm QPM}$ and 100 $W_0^{\rm QPM}$ vectors. Each $\xi_0^{\rm QPM}$ (and $W_0^{\rm QPM}$) vector has 46 values. The top plot shows distribution of values obtained 100 $\xi_0^{\rm QPM}$ vectors, and the bottom plot shows distribution of values gotten from 100 $W_0^{\rm QPM}$ vectors.}   
    \label{fig:QPM_Mono_Dist}     
\end{figure}

\begin{figure}
    \includegraphics[width=0.41\textwidth]{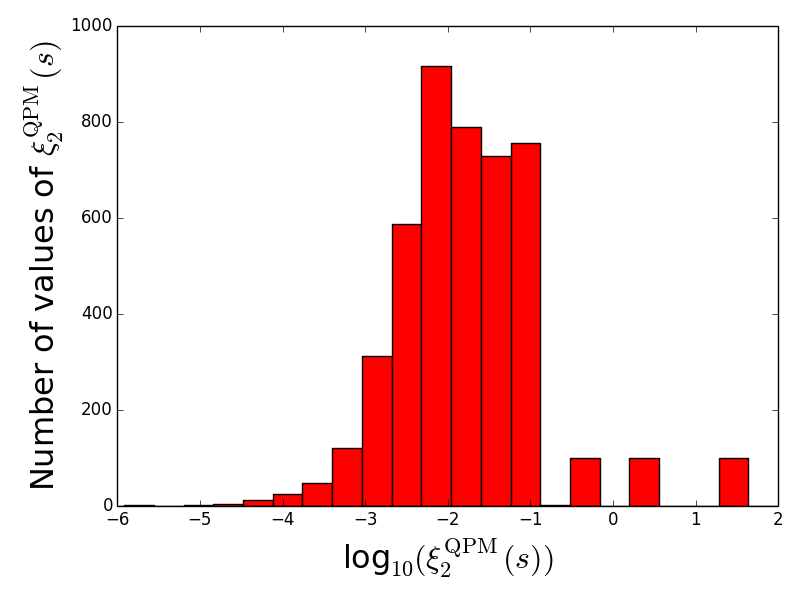}
    \includegraphics[width=0.41\textwidth]{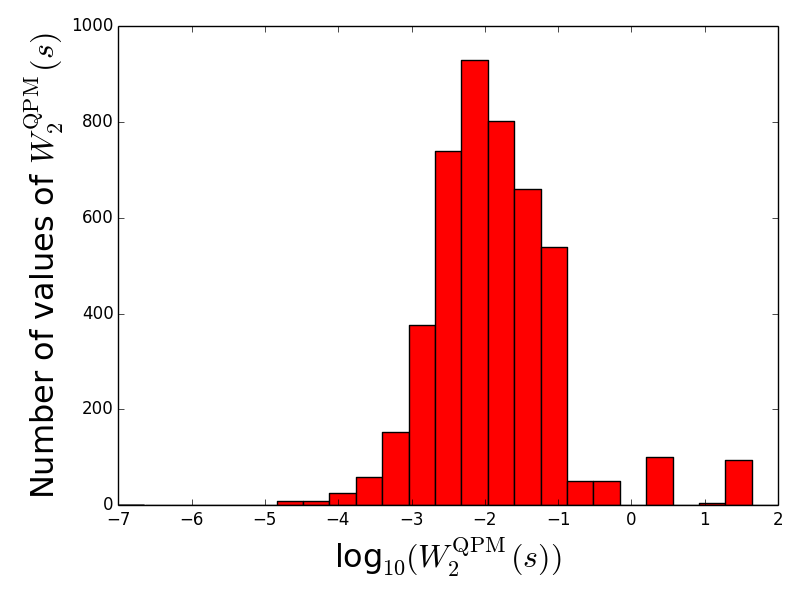}
    \caption{This figure shows the distribution of values of 100 $\xi_2^{\rm QPM}$ and 100 $W_2^{\rm QPM}$ vectors. Just like the $\xi_0^{\rm QPM}$ and $W_0^{\rm QPM}$ vectors, the $\xi_2^{\rm QPM}$ and $W_2^{\rm QPM}$ vectors also have 46 values each. The top plot displays distribution of values gotten from $\xi_2^{\rm QPM}$ vectors while the bottom plot displays the distribution of values obtained from $W_2^{\rm QPM}$ vectors.}   
    \label{fig:QPM_Quad_Dist}     
\end{figure}

\begin{figure}
    \includegraphics[width=0.41\textwidth]{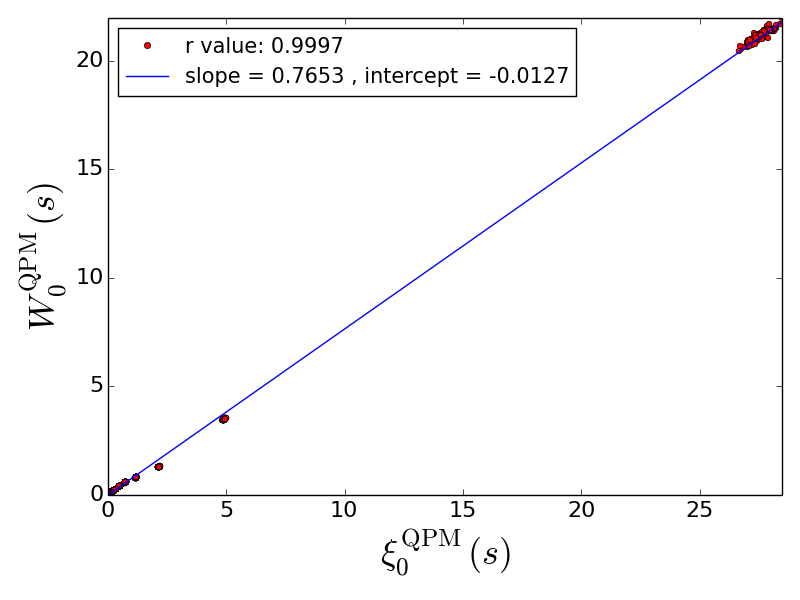}
    \includegraphics[width=0.41\textwidth]{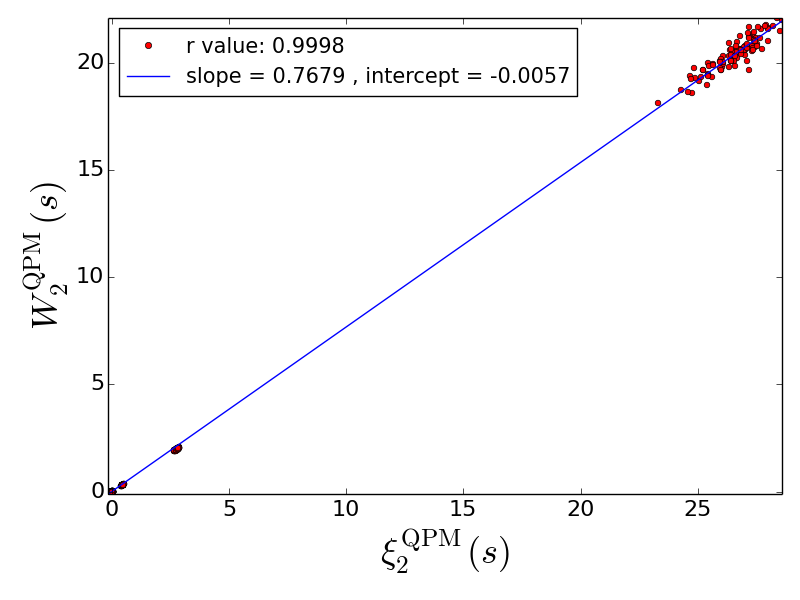}
    \caption{The top plot shows the correlation of points in 100 $\xi_{0}^{\rm QPM}$ vectors with the corresponding $W_{0}^{\rm QPM}$ vectors, while the bottom plot displays the correlation of points in 100 $\xi_{2}^{\rm QPM}$ vectors with the corresponding $W_{2}^{\rm QPM}$ vectors. Given the bin size and the range that we consider, each of these vectors has 46 numbers. We find that each $W$ vector obtained from QPM catalogue is very strongly correlated with its counterpart $\xi$ vector (gotten from QPM catalogue). In particular, we see that all numbers in $W_1$ vector are strongly correlated with $\xi_1$ vector, all numbers in $W_2$ vector are strongly correlated with $\xi_2$ vector and so on (all the way till $W_{100}$ and $\xi_{100}$, for both monopoles and quadrupoles). All in all, we have $46 \times 100 = 4600$ numbers from $W_0$ ($W_2$) and $4600$ counterparts in $\xi_0$ ($\xi_2$), and we find that there is a strong correlation between these two sets of 4600 points  (${\rm r value} \sim  0.9997$ for monopoles, and ${\rm r value} \sim 0.9998$ for quardupoles ). The blue lines in the top and the bottom plots represent the results of linear least-squares regression for measurements between points in $W$  and $\xi$ vectors.}    
    \label{fig:WXiCorrelation}
\end{figure}

The value of $\chi^2_{\mathcal{M}_0}/dof$ that we obtain in our comparison of $\mathcal{M}_0^{\rm theory}$  and $\mathcal{M}_0^{\rm obs}$, and the small error bars around $\mathcal{M}_0$ and $\mathcal{M}_2$ in Fig.~\ref{fig:LOWZ_Mark_ChosenParam} can be attributed to the nature of construction of the marked correlation function estimator ($\mathcal{M} = (1+W)/(1+\xi)$). At large distance scales, the values of marked weighted correlation functions ($W_{0,2}$) and standard correlation functions ($\xi_{0,2}$) are very small when compared to 1. Because of this, at large distance scales $\mathcal{M} = (1+W)/(1+\xi) \approx 1/1 = 1$ for all mocks. Figures~\ref{fig:QPM_Mono_Dist} and~\ref{fig:QPM_Quad_Dist} illustrate the distributions of values of $\xi_0^{\rm QPM}$, $W_0^{\rm QPM}$, $\xi_2^{\rm QPM}$ and $W_2^{\rm QPM}$ vectors obtained from QPM mocks. These histograms clearly show that most of the values in the monopoles and quadrupoles obtained from QPM mocks ($\xi_{0,2}^{\rm QPM}$, $W_{0,2}^{\rm QPM}$) are much lesser than one.

At small distance scales, the strong correlations between the mark-weighted correlation functions ($W_{0,2}$) and the standard correlated functions ($\xi_{0,2}$) play a major role in the small error bars that we see around $\mathcal{M}_{0,2}$. Fig.~\ref{fig:WXiCorrelation} shows the extent of correlation between values in $W_{0,2}$ and $\xi_{0,2}$ vectors obtained from QPM mocks. To get an intuitive idea of how the correlation between $W_{0,2}^{\rm QPM}$ and $\xi_{0,2}^{\rm QPM}$ affects the error bars around $\mathcal{M}_{0,2}$, let us take the help of a hypothetical new estimator say $N$, where $N_{0,2} = W_{0,2} / \xi_{0,2}$ . In this case it is clearly visible that when $W_{0,2}$ and $\xi_{0,2}$ are very strongly correlated, then $W_{0,2}/\xi_{0,2}$ will always be the same number(s) for all mocks. While the definition of the marked correlation function, $\mathcal{M}$ is not exactly equal to the hypothetical estimator $N$, it is not very far from it either ($N = W/\xi, \  \mathcal{M}=(1+W)/(1+\xi)$ ). Similarity in the constructions of $N$ and $\mathcal{M}$ would explain why we are seeing small error bars around $\mathcal{M}_0$ and $\mathcal{M}_2$ at small distance scales. At small distance scales, the strong correlations between the  $W_{0,2}$ and $\xi_{0,2}$ dominate (since the values of  $W_{0,2}$ and $\xi_{0,2}$ are comparatively larger on small distance scales.) Because of this, values of marked correlation function for different mocks are similar in small distance scales.

 So, it may not be very surprising why we are seeing a very small scatter around $\mathcal{M}_{0,2}$, but an expected amount of scatter around $\xi_{0,2}$ and $W_{0,2}$. However, the small scatter around the marked correlation function multipoles makes a $\chi^2/dof$ based comparison between LOWZ observation and GR simulation multipoles difficult. At the same time, one can make a robust comparison between observation and theory multipoles using an analysis based on fractional deviation (bottom panels of Fig.~\ref{fig:LOWZ_Mark_ChosenParam}).

\section{$\chi^2$ analysis in marked correlation functions using jackknife covariance matrices from \textsc{Elephant} GR simulation} \label{sec:chi2markedcorr_jn}

\begin{figure*}
\begin{multicols}{2}
    \includegraphics[width=0.42\textwidth]{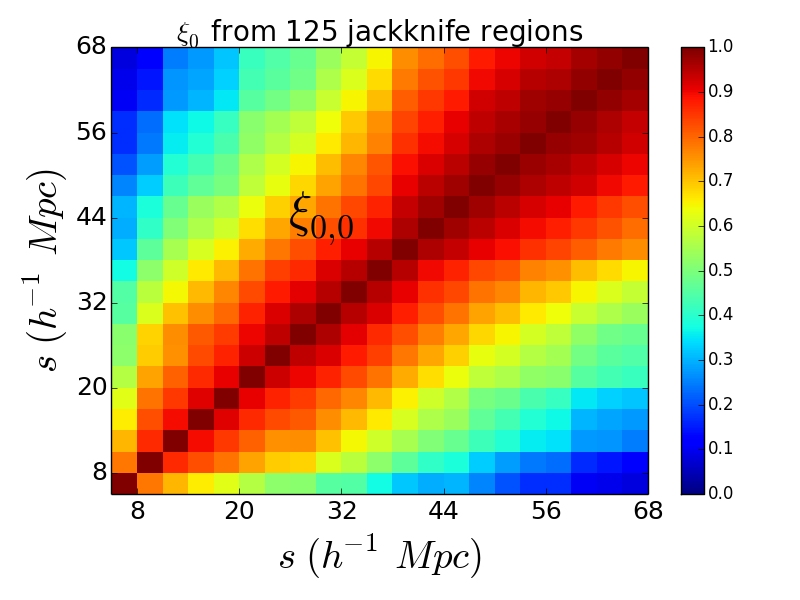}\par 
    \includegraphics[width=0.42\textwidth]{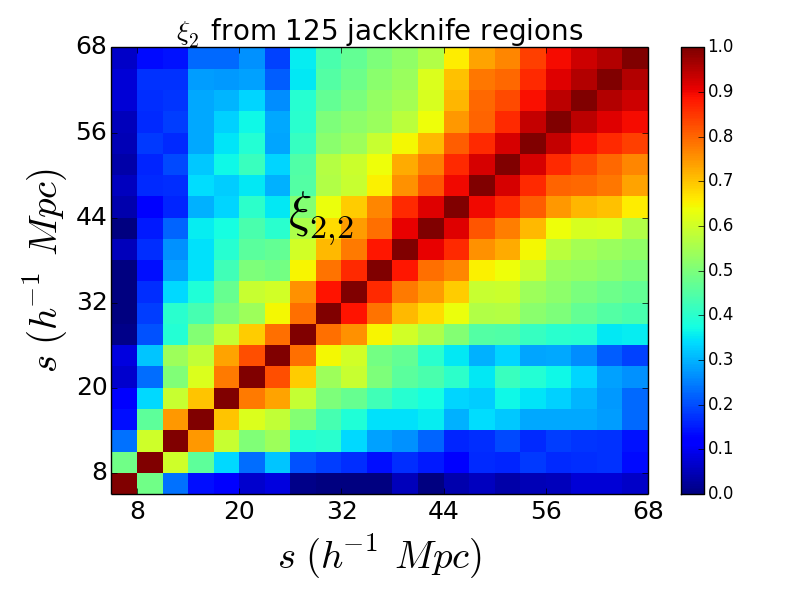}\par 
    \end{multicols}
\begin{multicols}{2}
    \includegraphics[width=0.42\textwidth]{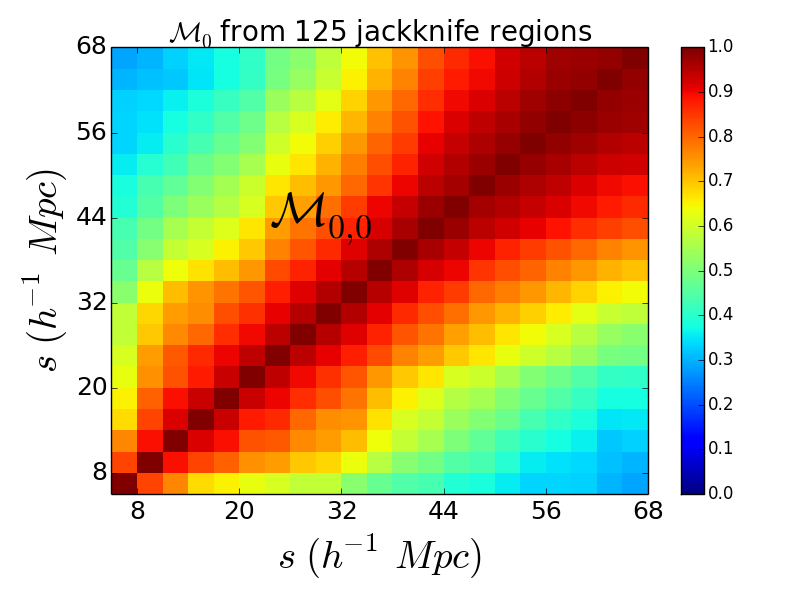}\par
    \includegraphics[width=0.42\textwidth]{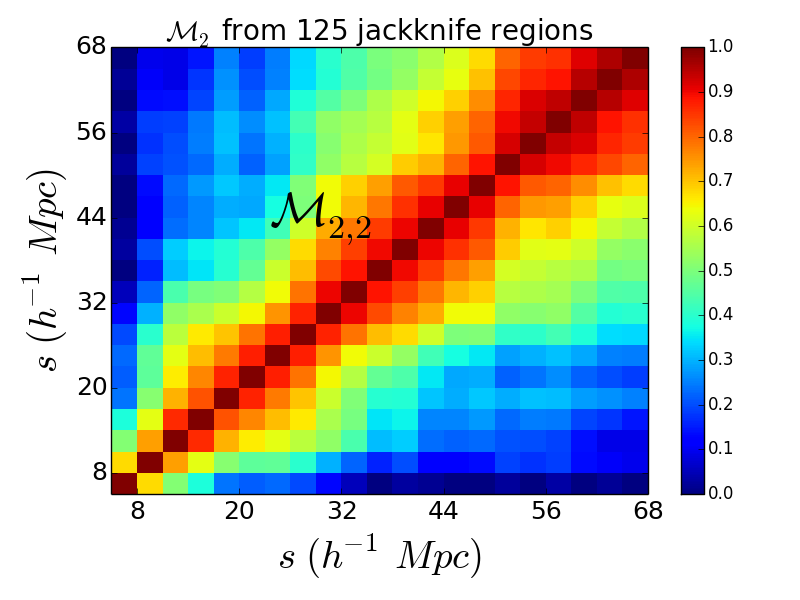}\par
\end{multicols}
\caption{The top plots show correlation matrices ($\mathbf{r}^{\xi}$) obtained from 125 regular correlation function monopoles (left) and 125 regular correlation function quadrupoles (right) which are gotten from jackknifing of \textsc{Elephant} GR simulation. The bottom plots show the correlation matrices ($\mathbf{r}^{ \mathcal{M} }$) for marked correlation function monopoles (left) and quadrupoles (right) which are obtained from the aforementioned 125 jackknife regions.}
\label{fig:JN_CorrMat}
\end{figure*}

In this section, we give results of correlation functions which are obtained from jackknifing of the \textsc{Elephant} GR simulation. We mask regions of size 204.8 $h^{-1}$Mpc (in $x$, $y$ and $z$ axes) in the GR simulation (total original volume $1024\times1024\times1024$ Mpc$^3$/h$^3$) to obtain 125 separate jackknife regions. Using these jackknife regions, one can obtain 125 separate two point correlation functions ($\xi_{0,2}^{\rm Jackknife}$), mark weighted correlation functions ($W_{0,2}^{\rm Jackknife}$) and marked correlation functions ($\mathcal{M}_{0,2}^{\rm Jackknife}$). One can use the 125 marked correlation function monopoles to construct covariance matrix $\Sigma_{\mathcal{M}_0}^{\rm Jackknife}$. Figure~\ref{fig:JN_CorrMat} shows correlation matrices ($\mathbf{r}_{ \xi_{0,2} }^{\rm Jackknife}$ and  $\mathbf{r}_{ \mathcal{M}_{0,2} }^{\rm Jackknife}$ ) obtained from these 125 jackknife regions.

Using $\chi^2_{\mathcal{M}_0}/dof$ analysis based on the above mentioned marked correlation monopole covariance matrix ($\Sigma_{\mathcal{M}_0}^{\rm Jackknife}$) and $\mathcal{M}_0^{\rm theory}$  and $\mathcal{M}_0^{\rm obs}$, one gets a value of $\chi^2_{\mathcal{M}_0}/dof = 2149$. This high value of  $\chi^2_{\mathcal{M}_0}/dof$ owes its origins to the same issues which affect $\chi^2_{\mathcal{M}_0}/dof$ analysis for QPM mocks. These issues have been discussed in detail in appendix~\ref{sec:chi2markedcorr}.

\begin{figure}
    \includegraphics[width=0.41\textwidth]{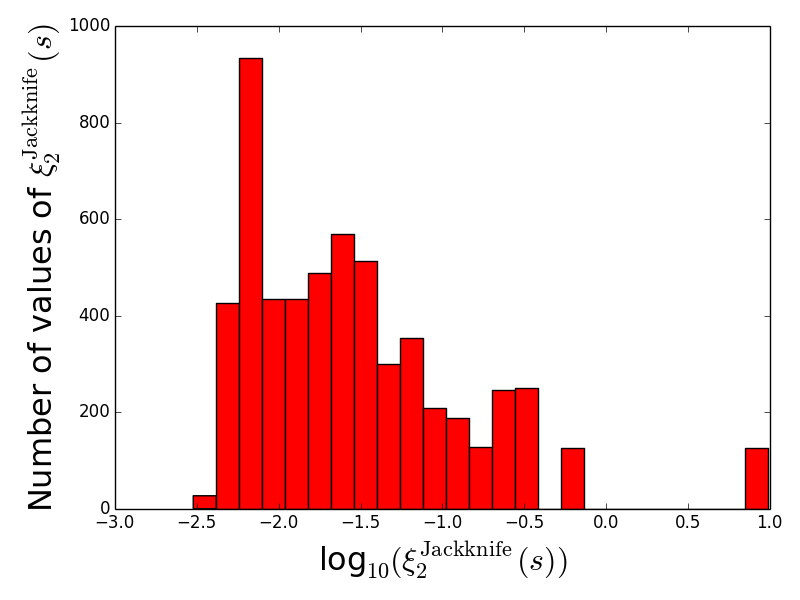}
    \includegraphics[width=0.41\textwidth]{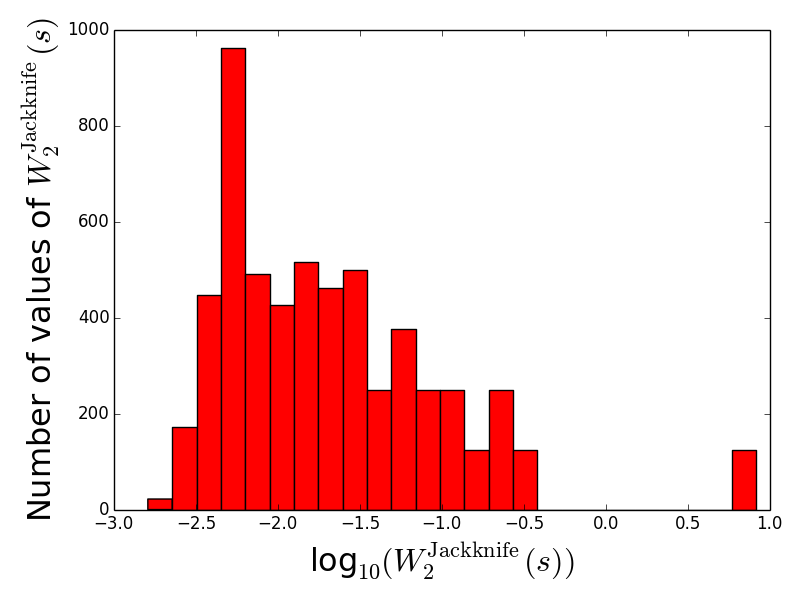}
    \caption{In this figure, we plot histogram of values of 125 $\xi_2^{\rm Jackknife}$ (top plot) and 125 $W_2^{\rm Jackknife}$ vectors (bottom plot). Like the $\xi_0^{\rm Jackknife}$ and $W_0^{\rm Jackknife}$ vectors, the $\xi_2^{\rm Jackknife}$ and $W_2^{\rm Jackknife}$ vectors also have 46 values each.}   
    \label{fig:JN_Quad_Dist}     
\end{figure}

\begin{figure}
    \includegraphics[width=0.41\textwidth]{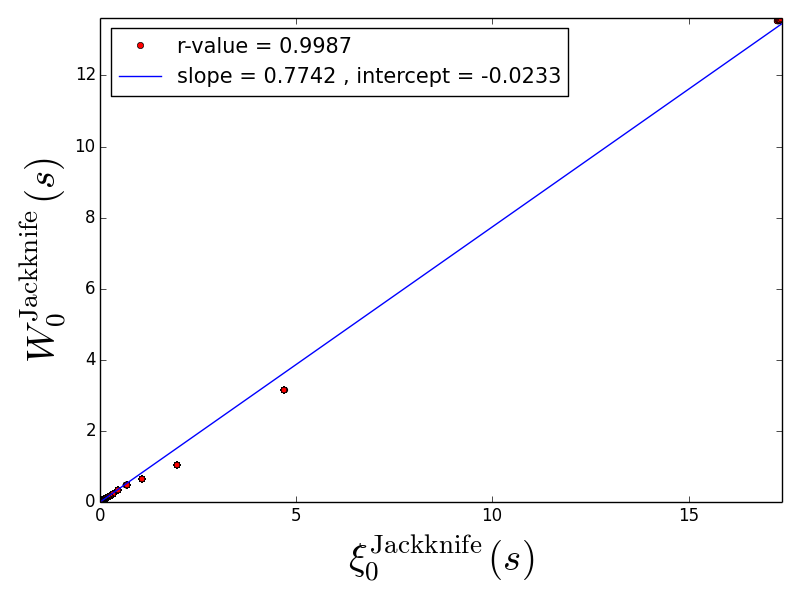}
    \includegraphics[width=0.41\textwidth]{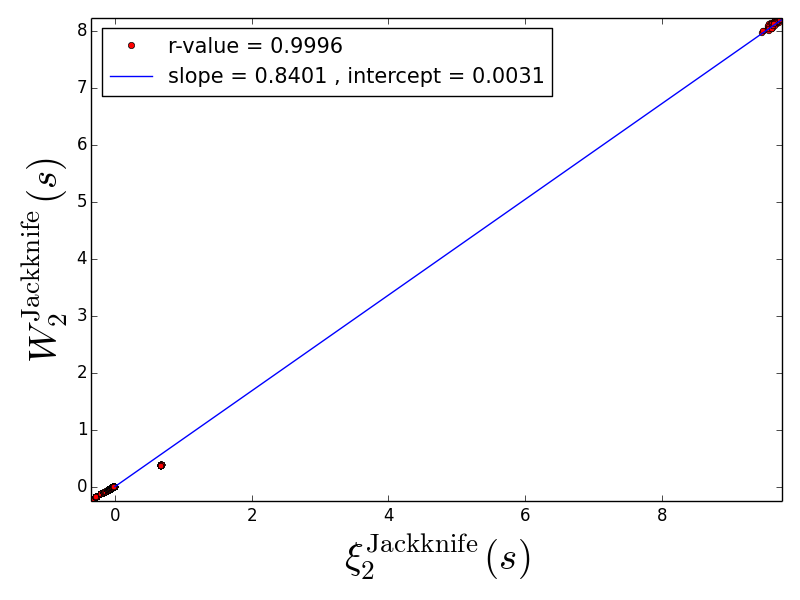}
    \caption{Here, we show the correlation of points in 125 $\xi_{0}^{\rm Jackknife}$ vectors with the corresponding $W_{0}^{\rm Jackknife}$ vectors in the top plot. The bottom plot displays correlation of points in 125 $\xi_{2}^{\rm Jackknife}$ vectors with points in the corresponding $W_{2}^{\rm Jackknife}$ vectors. Each of these vectors ($\xi_{0}^{\rm Jackknife}$, $W_{0}^{\rm Jackknife}$, $\xi_{2}^{\rm Jackknife}$, $W_{2}^{\rm Jackknife}$) has 46 numbers. The $W$ vectors are found to be very strongly correlated with their counterpart $\xi$ vectors. All in all, we have $46 \times 125 = 5750$ numbers from $W_0$ ($W_2$) and $5750$ counterparts in $\xi_0$ ($\xi_2$), and we find that there is a strong correlation between these two sets of 5750 points  (${\rm r value} \sim  0.9987$ for monopoles, and ${\rm r value} \sim 0.9996$ for quardupoles ). The blue lines in the top and the bottom plots represent the results of linear least-squares regression for measurements between points in $W$  and $\xi$ vectors.}    
    \label{fig:JN_WXiCorrelation}
\end{figure}

Figures~\ref{fig:JN_Mono_Dist} and~\ref{fig:JN_Quad_Dist} show histograms of values of multipoles ($\xi_{0,2}^{\rm Jackknife}$) and mark weighted mutipoles ($W_{0,2}^{\rm Jackknife}$) obtained from different jackknife regions. Comparison of these figures with figures~\ref{fig:QPM_Mono_Dist} and~\ref{fig:QPM_Quad_Dist} reveals that the reasons for high values of $\chi^2_{\mathcal{M}_0}/dof$ for covariance matrices obtained from QPM mocks and jackknife regions are similar. This is further supported by the strong correlations between $W_0^{\rm Jackknife}$---$\xi_{0}^{\rm Jackknife}$ vectors and $W_2^{\rm Jackknife}$ --- $\xi_{2}^{\rm Jackknife}$ which is illustrated in Figure~\ref{fig:JN_WXiCorrelation}.

\begin{figure}
    \includegraphics[width=0.41\textwidth]{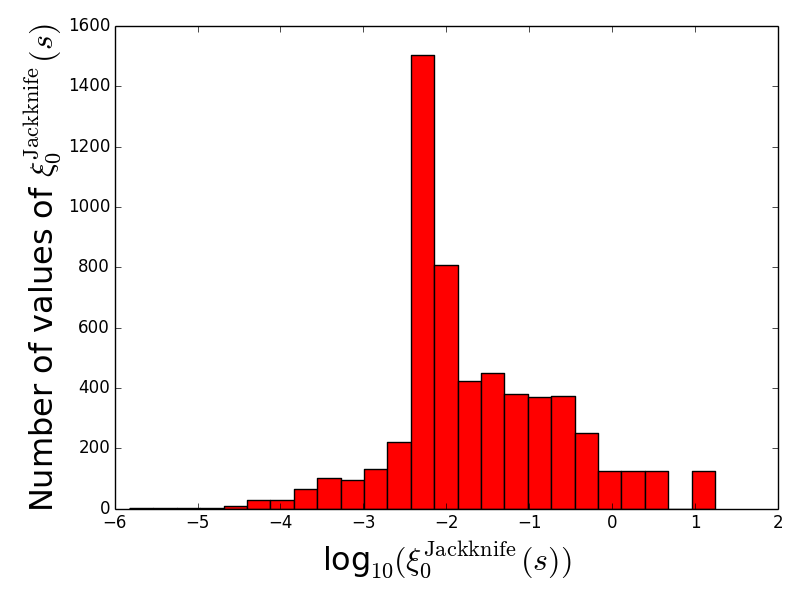}
    \includegraphics[width=0.41\textwidth]{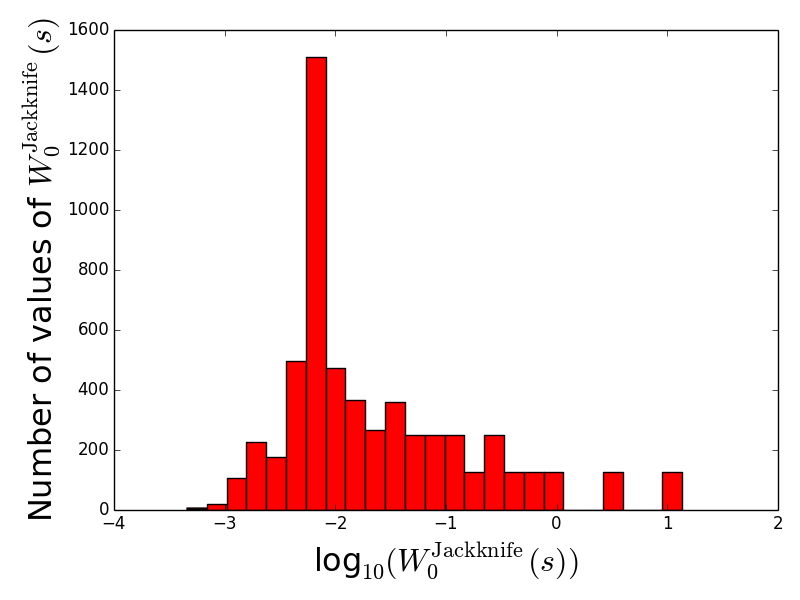}
    \caption{This figure shows distributions of values in 125 $\xi_0^{\rm Jackknife}$ (top plot) and 125 $W_0^{\rm Jackknife}$ vectors (bottom plot). Each $\xi_0^{\rm Jackknife}$ (and $W_0^{\rm Jackknife}$) vector has 46 values.}   
    \label{fig:JN_Mono_Dist}     
\end{figure}

\section{Standard and marked correlation functions}
\label{sec:MultipoleData}

Our LOWZ observational results are made available in tabulated form in Table~\ref{tab:MultipoleData}.

\begin{table*}
	\centering
	\caption{We give data for standard and marked correlation function multipoles (LOWZ observations and \textsc{elephant} GR simulations) in this table. }
	\label{tab:MultipoleData}
	\begin{tabular}{ccccccccc}
		\hline
		\hline
		$s$ & $\xi_0^{\rm obs}$ & $\xi_2^{\rm obs}$ & $\xi_0^{\rm theory}$ & $\xi_2^{\rm theory}$ & $\mathcal{M}_0^{\rm obs}$ & $\mathcal{M}_2^{\rm obs}$ & $\mathcal{M}_0^{\rm theory}$ & $\mathcal{M}_2^{\rm theory}$ \\
		\hline

		1.5 & $2.441 \times 10^{1}$ & $2.365 \times 10^{1}$ & $1.722 \times 10^{1}$ & $9.596 \times 10^{0}$ & $7.406 \times 10^{-1}$ & $7.608 \times 10^{-1}$ & $7.951 \times 10^{-1}$ & $8.581 \times 10^{-1}$\\	

		4.5 & $4.497 \times 10^{0}$ & $2.229 \times 10^{0}$ & $4.645 \times 10^{0}$ & $6.712 \times 10^{-1}$ & $6.803 \times 10^{-1}$ & $7.465 \times 10^{-1}$ & $7.281 \times 10^{-1}$ & $8.178 \times 10^{-1}$\\	

		7.5 & $1.937 \times 10^{0}$ & $2.700 \times 10^{-1}$ & $1.930 \times 10^{0}$ & $-3.061 \times 10^{-1}$ & $6.638 \times 10^{-1}$ & $8.895 \times 10^{-1}$ & $6.918 \times 10^{-1}$ & $1.084 \times 10^{0}$\\

		10.5 & $1.058 \times 10^{0}$ & $-1.298 \times 10^{-1}$ & $1.055 \times 10^{0}$ & $-3.384 \times 10^{-1}$ & $7.767 \times 10^{-1}$ & $1.016 \times 10^{0}$ & $7.942 \times 10^{-1}$ & $1.132 \times 10^{0}$\\	

		13.5 & $6.580 \times 10^{-1}$ & $-1.848 \times 10^{-1}$ & $6.591 \times 10^{-1}$ & $-2.683 \times 10^{-1}$ & $8.693 \times 10^{-1}$ & $1.047 \times 10^{0}$ & $8.847 \times 10^{-1}$ & $1.109 \times 10^{0}$\\	

		16.5 & $4.456 \times 10^{-1}$ & $-1.868 \times 10^{-1}$ & $4.471 \times 10^{-1}$ & $-2.029 \times 10^{-1}$ & $9.018 \times 10^{-1}$ & $1.064 \times 10^{0}$ & $9.122 \times 10^{-1}$ & $1.087 \times 10^{0}$\\	

		19.5 & $3.178 \times 10^{-1}$ & $-1.498 \times 10^{-1}$ & $3.175 \times 10^{-1}$ & $-1.617 \times 10^{-1}$ & $9.246 \times 10^{-1}$ & $1.052 \times 10^{0}$ & $9.301 \times 10^{-1}$ & $1.067 \times 10^{0}$\\	

		22.5 & $2.316 \times 10^{-1}$ & $-1.237 \times 10^{-1}$ & $2.394 \times 10^{-1}$ & $-1.371 \times 10^{-1}$ & $9.412 \times 10^{-1}$ & $1.046 \times 10^{0}$ & $9.451 \times 10^{-1}$ & $1.049 \times 10^{0}$\\	

		25.5 & $1.781 \times 10^{-1}$ & $-1.056 \times 10^{-1}$ & $1.832 \times 10^{-1}$ & $-1.048 \times 10^{-1}$ & $9.529 \times 10^{-1}$ & $1.040 \times 10^{0}$ & $9.559 \times 10^{-1}$ & $1.038 \times 10^{0}$\\															

		28.5 & $1.390 \times 10^{-1}$ & $-9.172 \times 10^{-2}$ & $1.435 \times 10^{-1}$ & $-9.287 \times 10^{-2}$ & $9.618 \times 10^{-1}$ & $1.033 \times 10^{0}$ & $9.636 \times 10^{-1}$ & $1.032 \times 10^{0}$\\															

		31.5 & $1.093 \times 10^{-1}$ & $-7.475 \times 10^{-2}$ & $1.123 \times 10^{-1}$ & $-7.523 \times 10^{-2}$ & $9.698 \times 10^{-1}$ & $1.025 \times 10^{0}$ & $9.708 \times 10^{-1}$ & $1.024 \times 10^{0}$\\															

		34.5 & $8.708 \times 10^{-2}$ & $-6.475 \times 10^{-2}$ & $9.027 \times 10^{-2}$ & $-7.013 \times 10^{-2}$ & $9.759 \times 10^{-1}$ & $1.023 \times 10^{0}$ & $9.758 \times 10^{-1}$ & $1.020 \times 10^{0}$\\															

		37.5 & $7.121 \times 10^{-2}$ & $-5.952 \times 10^{-2}$ & $7.377 \times 10^{-2}$ & $-5.838 \times 10^{-2}$ & $9.799 \times 10^{-1}$ & $1.021 \times 10^{0}$ & $9.803 \times 10^{-1}$ & $1.017 \times 10^{0}$\\															

		40.5 & $5.754 \times 10^{-2}$ & $-5.395 \times 10^{-2}$ & $5.928 \times 10^{-2}$ & $-5.267 \times 10^{-2}$ & $9.836 \times 10^{-1}$ & $1.017 \times 10^{0}$ & $9.836 \times 10^{-1}$ & $1.015 \times 10^{0}$\\															

		43.5 & $4.765 \times 10^{-2}$ & $-4.394 \times 10^{-2}$ & $4.837 \times 10^{-2}$ & $-4.162 \times 10^{-2}$ & $9.866 \times 10^{-1}$ & $1.017 \times 10^{0}$ & $9.870 \times 10^{-1}$ & $1.014 \times 10^{0}$\\															

		46.5 & $3.948 \times 10^{-2}$ & $-4.082 \times 10^{-2}$ & $3.820 \times 10^{-2}$ & $-3.976 \times 10^{-2}$ & $9.887 \times 10^{-1}$ & $1.013 \times 10^{0}$ & $9.885 \times 10^{-1}$ & $1.013 \times 10^{0}$\\															

		49.5 & $3.289 \times 10^{-2}$ & $-3.724 \times 10^{-2}$ & $3.224 \times 10^{-2}$ & $-3.691 \times 10^{-2}$ & $9.907 \times 10^{-1}$ & $1.013 \times 10^{0}$ & $9.906 \times 10^{-1}$ & $1.010 \times 10^{0}$\\															

		52.5 & $2.784 \times 10^{-2}$ & $-3.260 \times 10^{-2}$ & $2.672 \times 10^{-2}$ & $-3.568 \times 10^{-2}$ & $9.924 \times 10^{-1}$ & $1.011 \times 10^{0}$ & $9.924 \times 10^{-1}$ & $1.010 \times 10^{0}$\\															

		55.5 & $2.314 \times 10^{-2}$ & $-3.131 \times 10^{-2}$ & $2.104 \times 10^{-2}$ & $-3.138 \times 10^{-2}$ & $9.936 \times 10^{-1}$ & $1.010 \times 10^{0}$ & $9.942 \times 10^{-1}$ & $1.009 \times 10^{0}$\\		

		58.5 & $1.896 \times 10^{-2}$ & $-2.582 \times 10^{-2}$ & $1.620 \times 10^{-2}$ & $-2.734 \times 10^{-2}$ & $9.947 \times 10^{-1}$ & $1.008 \times 10^{0}$ & $9.952 \times 10^{-1}$ & $1.009 \times 10^{0}$\\																	

		61.5 & $1.592 \times 10^{-2}$ & $-2.196 \times 10^{-2}$ & $1.287 \times 10^{-2}$ & $-2.885 \times 10^{-2}$ & $9.957 \times 10^{-1}$ & $1.006 \times 10^{0}$ & $9.962 \times 10^{-1}$ & $1.008 \times 10^{0}$\\																	

		64.5 & $1.275 \times 10^{-2}$ & $-1.892 \times 10^{-2}$ & $1.045 \times 10^{-2}$ & $-2.711 \times 10^{-2}$ & $9.966 \times 10^{-1}$ & $1.005 \times 10^{0}$ & $9.972 \times 10^{-1}$ & $1.008 \times 10^{0}$\\																	

		67.5 & $1.104 \times 10^{-2}$ & $-1.738 \times 10^{-2}$ & $7.894 \times 10^{-3}$ & $-2.318 \times 10^{-2}$ & $9.971 \times 10^{-1}$ & $1.005 \times 10^{0}$ & $9.978 \times 10^{-1}$ & $1.008 \times 10^{0}$\\																	

		70.5 & $1.017 \times 10^{-2}$ & $-1.668 \times 10^{-2}$ & $7.721 \times 10^{-3}$ & $-2.243 \times 10^{-2}$ & $9.974 \times 10^{-1}$ & $1.005 \times 10^{0}$ & $9.983 \times 10^{-1}$ & $1.007 \times 10^{0}$\\																	

		73.5 & $9.174 \times 10^{-3}$ & $-1.510 \times 10^{-2}$ & $5.673 \times 10^{-3}$ & $-2.015 \times 10^{-2}$ & $9.978 \times 10^{-1}$ & $1.004 \times 10^{0}$ & $9.982 \times 10^{-1}$ & $1.006 \times 10^{0}$\\																	

		76.5 & $7.654 \times 10^{-3}$ & $-1.497 \times 10^{-2}$ & $5.044 \times 10^{-3}$ & $-1.820 \times 10^{-2}$ & $9.981 \times 10^{-1}$ & $1.004 \times 10^{0}$ & $9.986 \times 10^{-1}$ & $1.006 \times 10^{0}$\\																	

		79.5 & $6.100 \times 10^{-3}$ & $-1.347 \times 10^{-2}$ & $4.206 \times 10^{-3}$ & $-1.810 \times 10^{-2}$ & $9.987 \times 10^{-1}$ & $1.004 \times 10^{0}$ & $9.988 \times 10^{-1}$ & $1.005 \times 10^{0}$\\																	

		82.5 & $6.120 \times 10^{-3}$ & $-1.196 \times 10^{-2}$ & $3.308 \times 10^{-3}$ & $-1.519 \times 10^{-2}$ & $9.989 \times 10^{-1}$ & $1.003 \times 10^{0}$ & $9.990 \times 10^{-1}$ & $1.005 \times 10^{0}$\\																														

		85.5 & $5.200 \times 10^{-3}$ & $-6.878 \times 10^{-3}$ & $3.039 \times 10^{-3}$ & $-1.490 \times 10^{-2}$ & $9.989 \times 10^{-1}$ & $1.002 \times 10^{0}$ & $9.993 \times 10^{-1}$ & $1.005 \times 10^{0}$\\	

		88.5 & $5.280 \times 10^{-3}$ & $-6.091 \times 10^{-3}$ & $1.718 \times 10^{-3}$ & $-1.350 \times 10^{-2}$ & $9.989 \times 10^{-1}$ & $1.001 \times 10^{0}$ & $9.999 \times 10^{-1}$ & $1.004 \times 10^{0}$\\	

		91.5 & $4.989 \times 10^{-3}$ & $-4.801 \times 10^{-3}$ & $2.379 \times 10^{-3}$ & $-1.222 \times 10^{-2}$ & $9.988 \times 10^{-1}$ & $1.000 \times 10^{0}$ & $9.997 \times 10^{-1}$ & $1.005 \times 10^{0}$\\	

		94.5 & $6.273 \times 10^{-3}$ & $-3.058 \times 10^{-3}$ & $2.374 \times 10^{-3}$ & $-1.077 \times 10^{-2}$ & $9.988 \times 10^{-1}$ & $1.000 \times 10^{0}$ & $9.993 \times 10^{-1}$ & $1.003 \times 10^{0}$\\	

		97.5 & $5.711 \times 10^{-3}$ & $-3.590 \times 10^{-3}$ & $1.975 \times 10^{-3}$ & $-7.938 \times 10^{-3}$ & $9.988 \times 10^{-1}$ & $1.000 \times 10^{0}$ & $9.995 \times 10^{-1}$ & $1.002 \times 10^{0}$\\	

		100.5 & $5.284 \times 10^{-3}$ & $-2.643 \times 10^{-3}$ & $3.180 \times 10^{-3}$ & $-7.336 \times 10^{-3}$ & $9.990 \times 10^{-1}$ & $1.000 \times 10^{0}$ & $9.993 \times 10^{-1}$ & $1.002 \times 10^{0}$\\

		103.5 & $5.647 \times 10^{-3}$ & $-3.891 \times 10^{-3}$ & $3.713 \times 10^{-3}$ & $-5.841 \times 10^{-3}$ & $9.992 \times 10^{-1}$ & $1.000 \times 10^{0}$ & $9.992 \times 10^{-1}$ & $1.001 \times 10^{0}$\\	

		106.5 & $5.468 \times 10^{-3}$ & $-3.138 \times 10^{-3}$ & $3.689 \times 10^{-3}$ & $-5.890 \times 10^{-3}$ & $9.993 \times 10^{-1}$ & $1.000 \times 10^{0}$ & $9.992 \times 10^{-1}$ & $1.001 \times 10^{0}$\\

		109.5 & $4.786 \times 10^{-3}$ & $-2.859 \times 10^{-3}$ & $3.680 \times 10^{-3}$ & $-7.033 \times 10^{-3}$ & $9.993 \times 10^{-1}$ & $9.996 \times 10^{-1}$ & $9.994 \times 10^{-1}$ & $1.001 \times 10^{0}$\\

		112.5 & $4.780 \times 10^{-3}$ & $-2.944 \times 10^{-3}$ & $3.598 \times 10^{-3}$ & $-5.986 \times 10^{-3}$ & $9.993 \times 10^{-1}$ & $1.000 \times 10^{0}$ & $9.996 \times 10^{-1}$ & $1.001 \times 10^{0}$\\					

		115.5 & $3.465 \times 10^{-3}$ & $-3.965 \times 10^{-3}$ & $2.752 \times 10^{-3}$ & $-4.647 \times 10^{-3}$ & $9.995 \times 10^{-1}$ & $1.000 \times 10^{0}$ & $9.994 \times 10^{-1}$ & $1.001 \times 10^{0}$\\	

		118.5 & $2.977 \times 10^{-3}$ & $-3.590 \times 10^{-3}$ & $2.206 \times 10^{-3}$ & $-5.319 \times 10^{-3}$ & $9.999 \times 10^{-1}$ & $1.000 \times 10^{0}$ & $9.997 \times 10^{-1}$ & $1.002 \times 10^{0}$\\

		121.5 & $2.213 \times 10^{-3}$ & $-2.708 \times 10^{-3}$ & $1.190 \times 10^{-3}$ & $-5.262 \times 10^{-3}$ & $1.000 \times 10^{0}$ & $1.000 \times 10^{0}$ & $1.000 \times 10^{0}$ & $1.001 \times 10^{0}$\\					
																
		124.5 & $1.082 \times 10^{-3}$ & $-3.455 \times 10^{-3}$ & $4.723 \times 10^{-4}$ & $-7.014 \times 10^{-3}$ & $1.000 \times 10^{0}$ & $1.000 \times 10^{0}$ & $1.000 \times 10^{0}$ & $1.002 \times 10^{0}$\\		
																			
		127.5 & $1.351 \times 10^{-3}$ & $-3.506 \times 10^{-3}$ & $-6.736 \times 10^{-4}$ & $-8.059 \times 10^{-3}$ & $1.000 \times 10^{0}$ & $1.000 \times 10^{0}$ & $1.000 \times 10^{0}$ & $1.002 \times 10^{0}$\\																				

		130.5 & $1.648 \times 10^{-3}$ & $-3.537 \times 10^{-3}$ & $-1.443 \times 10^{-3}$ & $-7.672 \times 10^{-3}$ & $1.000 \times 10^{0}$ & $1.000 \times 10^{0}$ & $1.000 \times 10^{0}$ & $1.002 \times 10^{0}$\\	
																				
		133.5 & $1.884 \times 10^{-3}$ & $-3.672 \times 10^{-3}$ & $-2.086 \times 10^{-3}$ & $-8.335 \times 10^{-3}$ & $9.999 \times 10^{-1}$ & $1.000 \times 10^{0}$ & $1.000 \times 10^{0}$ & $1.002 \times 10^{0}$\\																							

		136.5 & $1.819 \times 10^{-3}$ & $-4.173 \times 10^{-3}$ & $-2.009 \times 10^{-3}$ & $-7.607 \times 10^{-3}$ & $1.000 \times 10^{0}$ & $9.998 \times 10^{-1}$ & $1.000 \times 10^{0}$ & $1.001 \times 10^{0}$\\

		\hline
	\end{tabular}
\end{table*}



\bsp	
\label{lastpage}
\end{document}